\documentclass[5p,times]{elsarticle}

\usepackage{graphicx}
\usepackage{amssymb,amsmath}
\usepackage{algorithm}
\usepackage{algpseudocode}
\usepackage{color}
\usepackage[table]{xcolor} 
\usepackage{xcolor} 
\usepackage{multirow} 
\usepackage[normalem]{ulem}
\usepackage{graphics}

\newcommand{\Input}{\textbf{Input: }}
\newcommand{\Output}{\textbf{Output: }}
\usepackage{hang} 
\newcommand{\R}{\mathbb{R}}
\DeclareMathOperator*{\argmin}{arg\,min}
\usepackage{physics} 

\newcommand\smallfont{\fontsize{10}{11}\selectfont}

\begin{document}


\title{A Study of Mixed Precision Strategies for GMRES on GPUs\\
}

\author{Jennifer A.\ Loe
, Christian A.\ Glusa
, Ichitaro Yamazaki
, Erik G.\ Boman
,\ and Sivasankaran Rajamanickam\fnref{a}}
\address{Center for Computing Research,  Sandia National Laboratories, Albuquerque, New Mexico, USA 87123\\
\{jloe, caglusa, iyamaza, egboman, srajama\}@sandia.gov}
\fntext[a]{Sandia National Laboratories is a multimission laboratory managed and operated by National Technology and Engineering Solutions of Sandia, LLC, a wholly owned subsidiary of Honeywell International, Inc., for the U.S. Department of Energy's National Nuclear Security Administration under contract DE-NA-0003525. This paper describes objective technical results and analysis. Any subjective views or opinions that might be expressed in the paper do not necessarily represent the views of the U.S. Department of Energy or the United States Government.  SAND2021-10892 O}

\begin{abstract}
Support for lower precision computation is becoming more common in accelerator hardware due to lower power usage, reduced data movement and increased computational performance. However, computational science and engineering (CSE) problems require double precision accuracy in several domains. This conflict between hardware trends and application needs has resulted in a need for mixed precision strategies at the linear algebra algorithms level if we want to exploit the hardware to its full potential while meeting the accuracy requirements. In this paper, we focus on preconditioned sparse iterative linear solvers, a key kernel in several CSE applications. We present a study of mixed precision strategies for accelerating this kernel on an NVIDIA V$100$ GPU with a Power 9 CPU. We seek the best methods for incorporating multiple precisions into the GMRES linear solver; these include iterative refinement and parallelizable preconditioners. Our work presents strategies to determine when mixed precision GMRES will be effective and to choose parameters for a mixed precision iterative refinement solver to achieve better performance. We use an implementation that is based on the Trilinos library and employs Kokkos Kernels for performance portability of linear algebra kernels. Performance results demonstrate the promise of mixed precision approaches and demonstrate even further improvements are possible by optimizing low-level kernels.
\end{abstract}

\maketitle


\section{Introduction}


In the current push towards exascale, modern supercomputers are increasingly relying on accelerator hardware for improved performance (with few exceptions). 
These accelerators are starting to support and even rely on lower precision computations as their primary use case. This is due to lower power usage, reduced data movement with lower memory footprint requirements, and increased computational performance for lower precision computations. The emergence of machine learning accelerators, such as Cerebras, Sambanova, and Graphcore, which support only lower precision, increases the adoption of lower precision even further. In addition to increased efficiencies, most of these accelerators are being designed to address the needs of machine learning use cases in the industry that can tolerate 32-bit or even 16-bit computations. 

Using lower precision is starting to become important to realize the full potential of emerging hardware.
However, computational science and engineering (CSE) problems have a need for 64-bit computations. This level of accuracy is important because several of these simulations are used for high-consequence decision making. 
This conflict between the hardware trend and the application requirements has resulted in a renewed interest in mixed precision algorithms at the linear algebra library level \cite{AnztMPOverview}.
Large-scale physics simulations with multiple discretized partial differential equations (PDEs) are also looking to take advantage of lower data precisions; however, unlike in machine learning, it is not obvious how to incorporate low precision data in the algorithm while obtaining double-precision accuracy of the final solution.

We focus on one of the expensive portions of solving PDEs, the sparse linear solve. While there are several approaches for solving sparse linear systems, we focus on sparse iterative linear solvers. The conjugate gradient (CG) method is highly effective for symmetric positive definite linear systems $Ax=b$. In this paper, we focus on the Generalized Minimum Residual method (GMRES) \cite{SaadSchultzGMRES}, which is commonly used for nonsymmetric systems. 

One algorithm that shows promise for this particular problem is GMRES with iterative refinement (GMRES-IR) \cite{WalkerTurnerGM-IR}. While the algorithm is several decades old, recent work with promising new analysis \cite{CarsonHighamNewAnalysis, CarsonIR3Precision} of this approach has increased interest. 
However, this method has not been well-studied on modern accelerator-based architectures, and the algorithm is not standard in linear solver software implementations. 
We address this gap by developing a Trilinos-based implementation of GMRES-IR. We further use this implementation for an experimental study that demonstrates the benefit of using GMRES-IR and, in some cases, what more needs to be improved. 

The main contributions of the paper are: 
\begin{itemize}
 \item Experimental evaluation of a Trilinos-based implementation of GMRES and two mixed precision variants, GMRES-IR and GMRES-FD (Float-to-Double), on GPUs; they show the promise of GMRES-IR for large problems that could take hundreds of iterations to converge.
 \item A demonstration with both model problems and general problems from the Suitesparse collection that GMRES-IR could reduce solve time by up to $1.5\times$ for preconditioned problems and $1.4\times$ for non-preconditioned problems while maintaining double precision accuracy. 
 \item An in-depth analysis of speedup of individual kernels within GMRES-IR on GPUs. 
 \item Evaluation of GMRES-IR combined with block Jacobi and polynomial preconditioning, and comparison with approaches such as low precision preconditioning with a higher precision solve.
 \item Evaluation of important GMRES-IR parameters such as subspace size, as well as suggestions for tuning them for best performance.
\end{itemize}

 Our aim is that these experimental results will help users to have realistic expectations about potential performance gains from GMRES-IR, a starting place for parameter selection, and an understanding of effective preconditioning choices. 
 
A preliminary version of this paper appeared in the proceedings of the AsHES workshop at IPDPS 2021 \cite{Loe2021experimental}. The current version includes more experiments, a more in-depth study of the SpMV performance, and an example that the right hand side may affect convergence and relative performance of the iterative methods.

\section{Related Work}

The strategy of using low-precision computations to obtain high-precision solutions goes back (at least) to the 1960s. Recently, there has been a renewed interest in mixed-precision (multiprecision) methods~\cite{AnztMPOverview, BaboulinDongarraMPAlgs}. 
The most successful approach for linear systems has been \emph{iterative refinement}~\cite{Moler-IR}. The key idea is to compute $A \approx LU$ in low precision, which is both faster and requires less memory than the standard double precision factorization. Initially, one solves for $Ax^0 \approx LU x^0 = b$ in low precision, but then computes the residual $r^k =b-Ax^k$ in high precision and solves for a correction term using the error equation $A \Delta x^k = r^k$. By updating the previous solution by the correction term, $x^{k+1} = x^k + \Delta x^k$, a more accurate solution is obtained. One can iterate (reusing the $LU$ factors) until the desired accuracy is reached, typically in just a few iterations.
Iterative refinement has been highly successful for dense systems, especially on GPUs~\cite{HaidarMPIterRefGPU,DongarraGPUIR}.

We focus on iterative methods for sparse systems, which do not require $LU$ factorization. Several recent works have studied using multiple precisions with GMRES, including \cite{ CarsonHighamNewAnalysis,CarsonIR3Precision, HaidarMPIterRefGPU,DongarraGPUIR,AnztMPIterRef,  GrattonExploitingVPGMRES,  LindquistGMRES, LindquistFullGIR, CarsonMultistage}. Anzt et al.\ \cite{AnztMPIterRef} analyzed iterative refinement combined with iterative solvers, viewing them as inner-outer solvers with iterative refinement as the outer solver and Krylov methods as the inner solver. They also presented some empirical results.

The original GMRES algorithm \cite{SaadSchultzGMRES} assumes every computation is done in high precision.
Turner and Walker \cite{WalkerTurnerGM-IR} observed that only a few key computations (including the residual) need to be done in high precision, while the rest can be done in lower precision. This approach has recently been revived as GMRES-IR \cite{CarsonHighamNewAnalysis,CarsonIR3Precision}.
A related approach is to compute the GMRES orthogonalization in lower (variable) precision~\cite{GrattonExploitingVPGMRES}.
Aliaga et al.~\cite{Aliaga-CB-GMRES} proposed to store the {GMRES} vectors in low precision but compute in high precision.
Another option is \emph{inexact Krylov} methods~\cite{SzyldInexactKrylov}, but this was designed for inexact matrix-vector products (only) and it is difficult to adapt to our mixed (single, double) precision use case.

The typical GMRES-IR implementations studied by Carson and Higham \cite{CarsonHighamNewAnalysis,CarsonIR3Precision} used various $LU$ factorizations in low precision as a preconditioner. There are two drawbacks of this approach. First, exact $LU$ may require too much memory due to fill (in the sparse case), has high computational complexity (due to fill), and may not be practical for large systems. Second, these preconditioners require a global triangular solve, which is not highly parallelizable, so not suitable for GPUs. Therefore, we do not consider $LU$-types of preconditioning here. 
The experiments in \cite{CarsonHighamNewAnalysis,CarsonIR3Precision} were limited to small problems in MATLAB on CPUs. 
Instead, we focus on classical sparse preconditioners such as block Jacobi and matrix polynomials, which are more efficient on GPUs. The GMRES-IR algorithm we consider is given in Algorithm \ref{alg:GMRES-IR} and is essentially the method by Turner and Walker \cite{WalkerTurnerGM-IR}.

Recently, Oktay and Carson \cite{CarsonMultistage}
have extended the GMRES-IR approach to a multistage setting. In practice, the theoretical analysis for GMRES-IR is quite pessimistic, so one may first try inexpensive computations and only switch to slower but more accurate precision when slow convergence (or divergence) is detected. Although our paper does not explicitly address this variation, we believe many of our insights could be useful in that approach. 

Also recently, in concurrent work, an empirical study of GMRES and GMRES-IR by Lindquist et al.\ \cite{LindquistGMRES,LindquistFullGIR} was presented. Although it is similar in scope, there are some differences. We study polynomial preconditioners, which are not considered in the Lindquist papers. We also provide kernel-level performance analysis (in particular, a model for speedup of SpMV), discuss the effect of different right-hand sides on convergence, and compare GMRES-IR to other schemes such as a precision switching scheme. \emph{The key contribution of this work is to evaluate this algorithm that shows promise in theory on a hardware that is designed to do well when using lower precision computation.} 

\section{GMRES and Mixed Precision Variants}

\begin{algorithm}[t]
  \caption{GMRES(m) (CGS) \cite[p.\ 172]{SaadItMeth} }
  \begin{hangingpar}
  \Input $A\in \R^{n\times n}$, $b \in \R^{n\times 1}$, initial guess $x_0\in R^{n\times 1}$, relative residual tolerance $rTol$
  \end{hangingpar}
  \Output approximate solution $x_m$
  \begin{algorithmic}[1]
    \State $r_0 = b-Ax_0$,
    \State $\gamma = \|{r_0}\|_2$, $v_1 = r_0/\gamma$, and $h_{1,1} = 0$
    \State Let $H_{:,j}$ be the vector of elements $\{h_{i,j}\}_{1\le i \le j}$.
      \For{$j=1:m$}
      \State $w_j = Av_j$
      \State Define $V_j = [v_1,v_2, \ldots, v_j]$. 
      \vspace{0.02in}
      \State $H_{:,j} = V_j^T w_j$ 
      \State $w_j = w_j - V_j H_{;,j}$ 
      \State $h_{j+1,j} = \|{w_j}\|_2$. (Lucky breakdown if $h_{j+1,j}=0$.)
      \State $v_{j+1}= w_j/h_{j+1,j}$
      \EndFor
    \State Define matrix $\overline{H}_m = \{h_{i,j}\}_{1\le i\le m+1, 1\le j \le m}$.
    \State Compute $\hat{d} = \argmin_{y\in \R^{m}} \|{\gamma e_1 - \overline{H}_m y}\|_2$, $\hat{x} = V_m\hat{d}$, and $x_m = x_0 + \hat{x}$.
    \State Compute $r_m = b - Ax_m$. If $\|{r_m}\|_2/\|{r_0}\|_2 \le rTol$, stop. Else, set $x_0 = x_m$, $r_0 = r_m$ and go to Step 2. 
  \end{algorithmic}
  \label{alg:GMRES(m)}
\end{algorithm}

We begin by describing GMRES, the computational kernels involved, and important observations for a mixed precision approach. We follow this with a description of two mixed precision variants, GMRES-IR and GMRES-FD.

\subsection{GMRES}
We consider real-valued $n\times n$ sparse linear systems $Ax=b$. GMRES(m) (Algorithm \ref{alg:GMRES(m)}) builds out a Krylov subspace $\mathcal{K}_m(A,b)=\text{span}\{b, Ab, A^2b, \ldots, A^{m-1}b\}$ from which to extract an approximate solution $\hat{x}$. 
At each iteration, GMRES appends a new basis vector to the subspace, orthogonalizes that vector against the previous basis vectors, and uses the expanded subspace to update the approximate solution $\hat{x}$. 

GMRES has ``converged" when the relative residual norm $\|b-A\hat{x}\|_2/\|b\|_2$ falls below some user-specified tolerance. We say that GMRES convergence has improved when either (a) the total solve time decreases or (b) the iteration count for convergence decreases. When computing with only one precision, (a) and (b) are roughly equivalent, but this will not always be the case when comparing double precision (fp64) GMRES with a mixed precision implementation. 

GMRES is optimal in the sense that it picks the approximate solution $\hat{x}$ so that the residual norm $\|b-A\hat{x}\|_2$ is minimized with $\hat{x} \in \mathcal{K}_m(A,b)$. When the dimension of the Krylov subspace becomes too large (i.e.\ orthogonalizing a new basis vector becomes too expensive or the set of $m$ basis vectors of length $n$ can no longer fit in memory), we \emph{restart} GMRES. This means that we discard the current Krylov subspace and start the GMRES iteration from the beginning with the new right-hand side $r=b-A(x_0-\hat{x})$. Then the final solution is the sum of the initial starting vector $x_0$ and all intermediate solution vectors $\hat{x}$. We refer to the value $m$ as the \textit{maximum subspace size} or the \textit{restart length} for GMRES.

Note that restarting GMRES can slow convergence. When restarted, GMRES loses crucial eigenvector information from the previous subspace that allows it to converge more quickly to a solution \cite{MorganGMRESE}. It has to recreate this information in the next subspace, which requires more time and iterations. Thus, it can be a challenge to choose a restart length for GMRES that is large enough for quick convergence but small enough to fit in memory on GPU accelerators. 

The primary sources of computational expense for GMRES are (1) sparse matrix-vector products (SpMVs) with the matrix $A$ (Alg.\ \ref{alg:GMRES(m)}, line $5$) and (2) orthogonalization of the Krylov subspace vectors. In our experiments, each GMRES iteration uses two passes of classical Gram-Schmidt orthogonalization (CGS2). Each of these two orthogonalization passes requires two calls to GEMV, one with a transpose to compute inner products, and another with no transpose to subtract out components of the previous vectors (Alg.\ \ref{alg:GMRES(m)}, lines $7$ and $8$). Other less expensive operations include norms, small dense matrix operations with matrix $H$, and vector additions. 

\subsection{Mixed Precision GMRES-IR}
For GMRES with iterative refinement (GMRES-IR), we will run the GMRES algorithm in \emph{single precision} (fp32) and then ``refine" the algorithm at each restart by starting the next GMRES run with a right-hand-side vector that has been computed in \emph{double precision} (fp64). See Algorithm \ref{alg:GMRES-IR} for more details. We maintain both double and single precision copies of the matrix $A$ in memory for performing SpMVs in the appropriate precision. 
Note that we only check for convergence of GMRES-IR at each restart, when the residuals are recomputed. This is different from standard GMRES where we can monitor an implicit residual within the iteration to alert us to convergence. This is a less-than-ideal implementation, but it allows us to work within the Belos software structure (see Section \ref{sec:software}). Since we give a new right-hand side vector to the inner Belos solver at each restart, Belos is unable to monitor the convergence of the original (outer) problem.   
Thus, for this implementation, GMRES-IR may take at most $m-1$ extra iterations in single precision over what is absolutely needed for convergence. We include cost of such iterations in our performance comparisons.  Future Belos implementations of GMRES-IR will have mechanisms to overcome this limitation. 
\begin{algorithm}
\caption{GMRES-IR}\label{alg:GMRES-IR}
  \begin{algorithmic}[1]
  \State $r_0 = b-Ax_0$ [double]
  \For{ $i=1,2, \ldots$ until convergence:}
    \State GMRES$(m)$ solves $Au_i = r_i$ for correction $u_i$ [single]
    \State $x_{i+1} = x_i + u_i$ [double]
    \State $r_{i+1} = b - Ax_{i+1}$ [double]
  \EndFor
  \end{algorithmic}
\end{algorithm}

\subsection{Mixed Precision GMRES-FD}

A first inclination when attempting to incorporate low precision into GMRES$(m)$ is to perform the entire first part of the calculation in one precision and then switch precisions at one of the restarts. We briefly explore these possibilities and then demonstrate why GMRES-IR is the better candidate for incorporating low precision. 

There are two options for switching precisions mid-solve: (1) Start in single precision and later switch to double, or (2) start in double and switch to single precision. The theory of \emph{inexact Krylov} supports option (2), stating that one can loosen the accuracy of the matrix-vector multiply (SpMV) as the iteration progresses and still converge to the correct solution \cite{SzyldInexactKrylov}. Furthermore, \cite{GrattonExploitingVPGMRES} shows that one can loosen the accuracy of inner products in addition to accuracy of the SpMV and still get convergence behavior close to that of full double precision. 
Option (2) may also be preferable because the initial computations in double precision may allow the Krylov subspace to quickly get good approximations to key eigenvectors, which can aid convergence \cite{MorganGMRESE}. However, inexact Krylov theory assumes that the vector operations are done in full precision and only the matrix-vector multiply is inexact. Therefore, the theory does not cover the use case of switching from double to single precision. 
It is not clear if a single precision solver can even converge to double precision accuracy; thus, we do not evaluate option (2) in our experiments. We assess option (1), switching from single precision GMRES$(50)$ to double precision GMRES$(50)$ and using the single precision solution vector as a starting vector for the double precision GMRES iteration. We call this method GMRES-FD (Float-Double).

\subsection{Preconditioning} We investigate two lower precision alternatives to traditional fp64 preconditioning: (a) double precision GMRES with a single precision preconditioner and (b) GMRES-IR with a single precision preconditioner. Most previous studies of GMRES-IR (e.g.\ \cite{CarsonIR3Precision}) used some variation on LU preconditioning. 
Here, we investigate more paralellizable preconditioners, using a polynomial preconditioner (Section \ref{subsec:precon_compar}) and a block Jacobi preconditioner (Section \ref{subsec:suitesparse}). 
In all tests, we use right preconditioning ($AMM^{-1}x = b$) so that the residuals of the preconditioned problem match those of the unpreconditioned problem in exact arithmetic. Each time an fp32 preconditioner $M$ is applied to an fp64 vector $x$ in case (a), we must cast $x$ to fp32, multiply it by $M$ in fp32, and cast the result back to fp64. For case (b), $M$ is both computed and applied entirely in fp32. 
 Polynomial preconditioning is applied as follows: We use a polynomial preconditioner based upon the GMRES polynomial (see details in \cite{LoePPTrilinos}). Here, using a polynomial $p(t)=\sum_{k=0}^{d}c_kt^k$ of degree $d$ as a preconditioner $M$ is to be understood as $M = p(A) = \sum_{k=0}^{d}c_k A^{k}$. (See \cite{XiaoPPCG} for a related study with the conjugate gradient method run in double precision with a single precision polynomial preconditioner.) 

\section{Software Implementation}
\label{sec:software}
Trilinos \cite{Trilinos} is a large software library with packages for PDE discretizations, linear and non-linear solvers, preconditioners, partitioners, and distributed linear algebra. We use the Trilinos framework for our solver implementation, with the eventual goal of making GMRES-IR available in the public codebase. Thus, we test GMRES and GMRES-IR within the framework of Belos \cite{ThornquistBelosAmesos2}, the Trilinos sparse iterative linear solvers package. Our final software version will be available with Tpetra-based linear algebra and will run with MPI over many CPUs and GPUs. For this paper, however, we only consider solvers on a single CPU/GPU. For the solvers' linear algebra backend, we use the Kokkos \cite{TrottKokkos3, TrottKokkosEco} and Kokkos Kernels \cite{RajamanickamKokkosKernels} libraries, which provide portable, optimized linear algebra operations for GPUs. 


 The Belos linear solvers package does not contain its own implementation of linear algebra, but instead relies on abstracted linear algebra interfaces through the \texttt{Belos::MultiVectorTraits}. We created a Kokkos-based adapter for Belos, letting the Kokkos adapter inherit from \texttt{Belos::MultiVector}. All of the length $n$ basis vectors for the Krylov subspace are stored in \texttt{Kokkos::Views} and operated on via the MultiVector interface. The interface implements all the needed capabilities to solve linear systems $Ax=b$ with a single right-hand side. 
Belos' solvers are all templated upon a user-specified scalar type, so they can be run in either float or double precision. Thus, at first glance, it seems that they would be well-suited for mixed precision computations. However, these templates assume that all operations are carried out in the same scalar type; there are no current capabilities to mix and match precisions within a solver. In spite of this, it is possible to perform operations outside of a solver using a different precision from the one the solver uses. We do this in our GMRES-IR implementation: The code initializes a Belos GMRES solver in fp32. At each restart, we retrieve the current solution vector from the Belos solver and convert it to fp64. Then we compute the current residual, convert that residual vector back to fp32, and feed that residual to the fp32 GMRES solver as the next right-hand side. 

\textit{Limitations of current implementation:}
Since we use the existing Belos interface, any mixed precision operations that are internal to the solver must be handled entirely in the linear algebra adapter. 
In order to avoid this difficulty, we do not study variations of GMRES where internal kernels use lower precision, e.g.\ GMRES with mixed precision orthogonalization or low precision SpMVs.

Additionally, the Belos linear solvers package was not designed with GPUs or other accelerators in mind: Belos requires that results of some GPU operations be stored in a dense matrix representation on host (\texttt{Teuchos::SerialDenseMatrix}). 
This requires data movement between the GPU and CPU along with memory allocations that otherwise might be unnecessary. 
Furthermore, the structure in Belos forces separate kernel launches for each GPU operation, while in a Kokkos-only implementation some of these operations could be fused. We plan to improve upon these limitations in future software upgrades of the Belos package.

\section{Experimental Results}
\label{sec:experiments}


All experiments that follow are run on a node equipped with a Power 9 CPU that has 318 GB DDR3 RAM and a Tesla V$100$ GPU with $16$ GB GDDR5 RAM. 
We used GCC 7.2.0, CUDA 9.2.88, Kokkos and Kokkos Kernels 3.2.0 and Trilinos 13.1. 
All PDE test problems either come from the SuiteSparse Matrix Collection \cite{DavisSparseCollect} or were generated with finite difference stencils via the Trilinos Galeri package.

The following experiments are run as follows: Unless otherwise stated, we restart both double precision GMRES$(m)$ and GMRES-IR after each run of $m=50$ iterations. 
All solvers are run to a relative residual convergence tolerance of $1\mathrm{e}{-}10$.
For each problem, we use a right-hand side vector $b$ of all ones and a starting vector $x_0$ of all zeros. For each set of results, we exemplify the run that has the median of three solve times.
For GMRES-IR , total solve times do not include the time needed to make a single precision copy of matrix $A$, but they do include time required to convert residual vectors from double to single precision (and vice-versa) during the refinement stage. Note that results are not entirely deterministic; numerical errors from reductions on the GPU can give slightly different convergence behaviors. 

The rest of the experiments section is organized as follows. We compare different approaches for mixed precision GMRES (\ref{subsec:IRvsFD}). We evaluate GMRES and mixed precision GMRES-IR unpreconditioned (\ref{subsec:unprecon_compar}) and preconditioned (\ref{subsec:precon_compar}) for their convergence and performance. We also examine the effect of the right-hand side vector on convergence (\ref{sec:rhscomp}). Next we do an in-depth analysis of the performance we observe in SpMV (\ref{sec:SpMV} and \ref{sec:SpMVTestSet}). We also study how choice of GMRES restart size affects performance (\ref{subsec:restart}). Finally, we evaluate our approach on a few general problems from the SuiteSparse collection (\ref{subsec:suitesparse}). 

\subsection{GMRES-IR vs GMRES-FD}
\label{subsec:IRvsFD}

\begin{figure}
  \centering
  \includegraphics[width=0.4\textwidth,keepaspectratio]{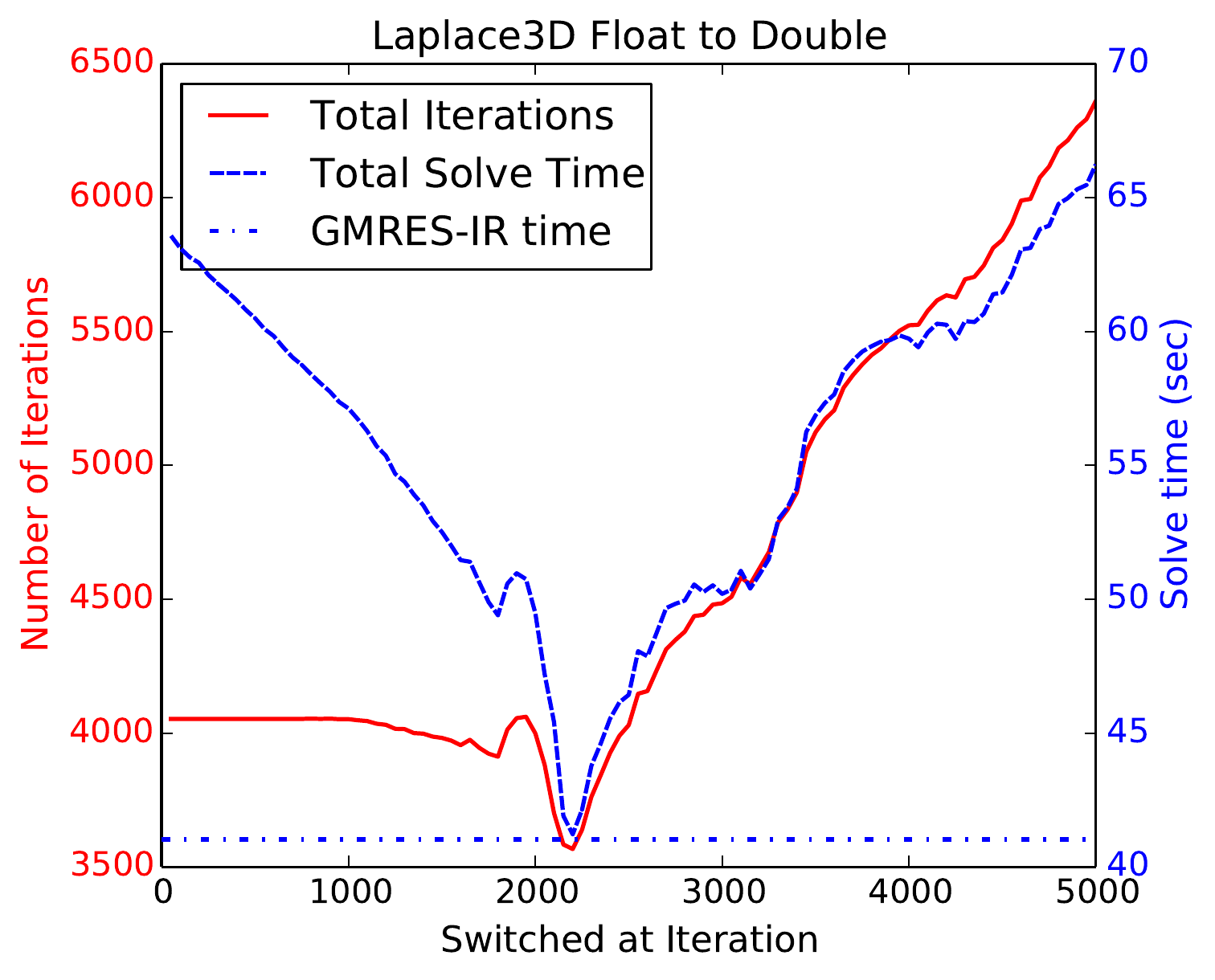}
  \caption{Total solve time and iteration count for a 3D Laplacian with GMRES-FD, switching from single precision to double precision at the iteration indicated on the horizontal axis. Dotted line at bottom indicates solve time for GMRES(50)-IR.}
  \label{fig:LaplSwitchIter}
\end{figure}

\begin{figure}
  \centering
  \includegraphics[width=0.4\textwidth,keepaspectratio]{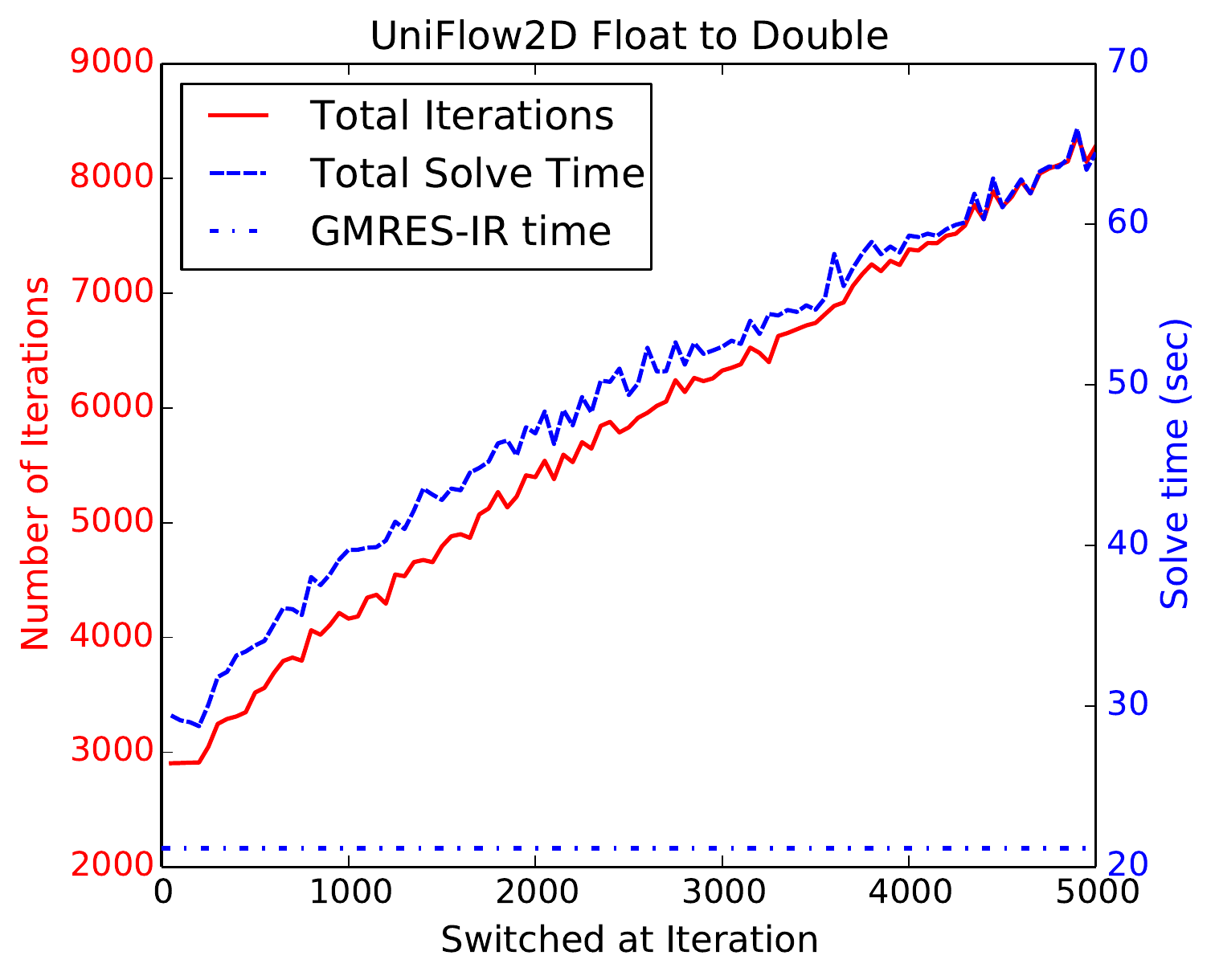}
  \caption{Total solve time and iteration count for the problem UniFlow2D2500 with GMRES-FD, switching from fp32 to fp64 at the iteration indicated on the horizontal axis. Dotted line at bottom indicates solve time for GMRES(50)-IR.}
  \label{fig:UniFlowSwIter}
\end{figure}

We begin experimental evaluations by comparing GMRES(m) in double precision, GMRES-IR, and GMRES-FD.
The first question with GMRES-FD: At what point is the right moment to switch precisions? We investigate with two different problems, comparing \emph{multiple runs of GMRES-FD} (switching at different iteration numbers), with a single run of GMRES-IR and GMRES(m). The first problem is a Laplacian from a 3D finite difference stencil with grid size $200$, and the second is a 2D convection-diffusion problem named ``UniFlow" with grid size $2500$. For both of these problems, we tested GMRES-FD, switching from fp32 to fp64 at each multiple of $50$ iterations (so at each restart). The $x$-axis in Figures \ref{fig:LaplSwitchIter} and \ref{fig:UniFlowSwIter} indicates the iteration at which the solver switched from float to double precision. The left vertical axis gives the total number of iterations required for convergence (the sum of single and double precision iterations). The right vertical axis gives total solve time for the problem. 

One can predict that switching to fp64 too early is not harmful to convergence, but it does not take full advantage of the fp32 solver to find the minimum solve time. 
If the chosen switching point is too late, then the fp32 solver takes extra iterations, adding to the total solve time but not making any progress. This is exactly what we see with the Laplacian problem in Figure \ref{fig:LaplSwitchIter}. The solve time slowly for GMRES-FD decreases until reaching a minimum when the switch happens at $2200$ iterations. Here, the total number of iterations required for the solve is $3567$, while the solve time is $41.22$ seconds. 
Comparatively, GMRES$(50)$-IR converges in $4100$
iterations and $41.03$ seconds. The double precision-only problem requires $4053$ iterations and $63.83$ seconds. Thus, GMRES-IR attains the minimum solve time of all methods without needing to manually determine when to switch precisions. 
Results from testing various switching points for GMRES-FD on the UniFlow problem (Figure \ref{fig:UniFlowSwIter}) are somewhat counterintuitive. The minimum of $28.77$ seconds (with a total of $2911$ iterations) occurs when switching at only $200$ iterations. This gives little improvement over the purely double precision solver, which required $2905$ iterations and $29.62$ seconds. Did the single precision solver's convergence stall after only $200$ iterations? Not at all! At a switching point of $2800$ iterations, for instance, the initial vector $x_0$ from the fp32 solver helps the fp64 solver to start with an initial residual norm of $9.9\mathrm{e}{-}5$. However, even with the good starting vector, the fp64 solver still needs an additional $3295$ iterations to converge. We hypothesize that this is because the new $x_0$ used at the switch of precisions did not contain eigenvector components that were present in the original right-hand side $b$. 
GMRES-IR, on the other hand, converges in $3000$ iterations and only $21.17$ seconds. It is the best method by far. This experiment demonstrates a case where GMRES-IR is quite helpful and GMRES-FD is mostly ineffective. 
We will use GMRES-IR as the mixed precision approach for the rest of the paper.
Next we look at how convergence of GMRES-IR compares to GMRES double and which kernels contribute most to speedup. 



\subsection{Convergence and Kernel Speedup for GMRES vs GMRES-IR}
\label{subsec:unprecon_compar}
\begin{figure}
  \centering
  \includegraphics[width=0.45\textwidth,keepaspectratio]{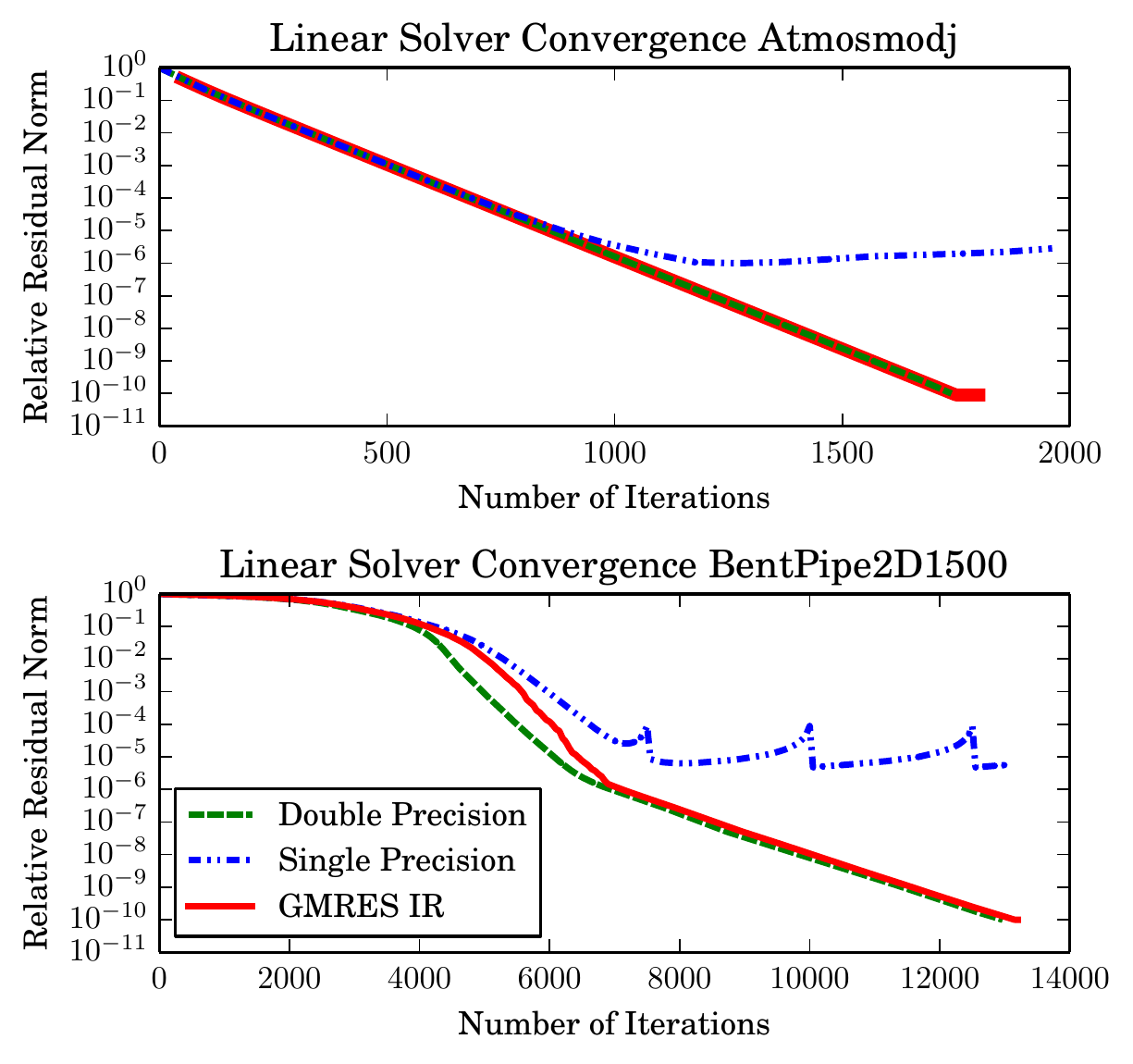}
  \caption{Relative residual norm convergence for matrix atmosmodj (top) and BentPipe2D1500 (bottom). Single precision GMRES$(50)$ is represented by the blue dash-dot line, double precision by green dashes, and mixed precision GMRES$(50)$-IR by the red solid line.}
  \label{fig:BentPipeConv}
\end{figure}

We next consider matrices BentPipe2D1500 and atmosmodj.  BentPipe2D1500 is a 2D convection-diffusion problem with $nx=1500$, $n=2{,}250{,}000$ and $nnz = 11{,}244{,}000$. (Here $nx$ denotes the number of grid points in each direction of the mesh for the finite difference discretization of the PDE, and $nnz$ denotes the number of nonzero elements in the sparse matrix $A$.) The underlying PDE is strongly convection-dominated, so the matrix is ill-conditioned and highly non-symmetric. The problem atmosmodj, taken from the SuiteSparse \cite{DavisSparseCollect} matrix collection, represents a computational fluid dynamics problem from atmospheric modeling.  Its size is $n=1{,}270{,}432$ with $nnz = 8{,}814{,}880$.
We compare GMRES$(50)$ in all single precision, GMRES$(50)$ in all double precision, and GMRES$(50)$-IR. Convergence plots are in Figure \ref{fig:BentPipeConv}. 
For both problems, the fp32 solver reaches a minimum relative residual norm near $1\mathrm{e}{-}6$. To converge to a tolerance of $1\mathrm{e}{-}10$, the fp64 GMRES solver needs $1740$ iterations for atmosmodj and $12{,}967$ iterations for BentPipe2D1500. GMRES-IR converges in $1750$ iterations for atmosmodj and $13{,}150$ iterations for BentPipe2D1500.  
Observe that for both problems, the convergence curve of GMRES-IR closely follows that of the double precision solver. This phenomenon is related to the theory built by \cite{GrattonExploitingVPGMRES} for non-restarted GMRES; it has also been observed by \cite{LindquistGMRES} for restarted GMRES.
To reiterate, \emph{the convergence of the mixed precision version of the solver follows the double precision version closely}. 

\begin{figure}
  \centering
  \includegraphics[width=0.45\textwidth,keepaspectratio]{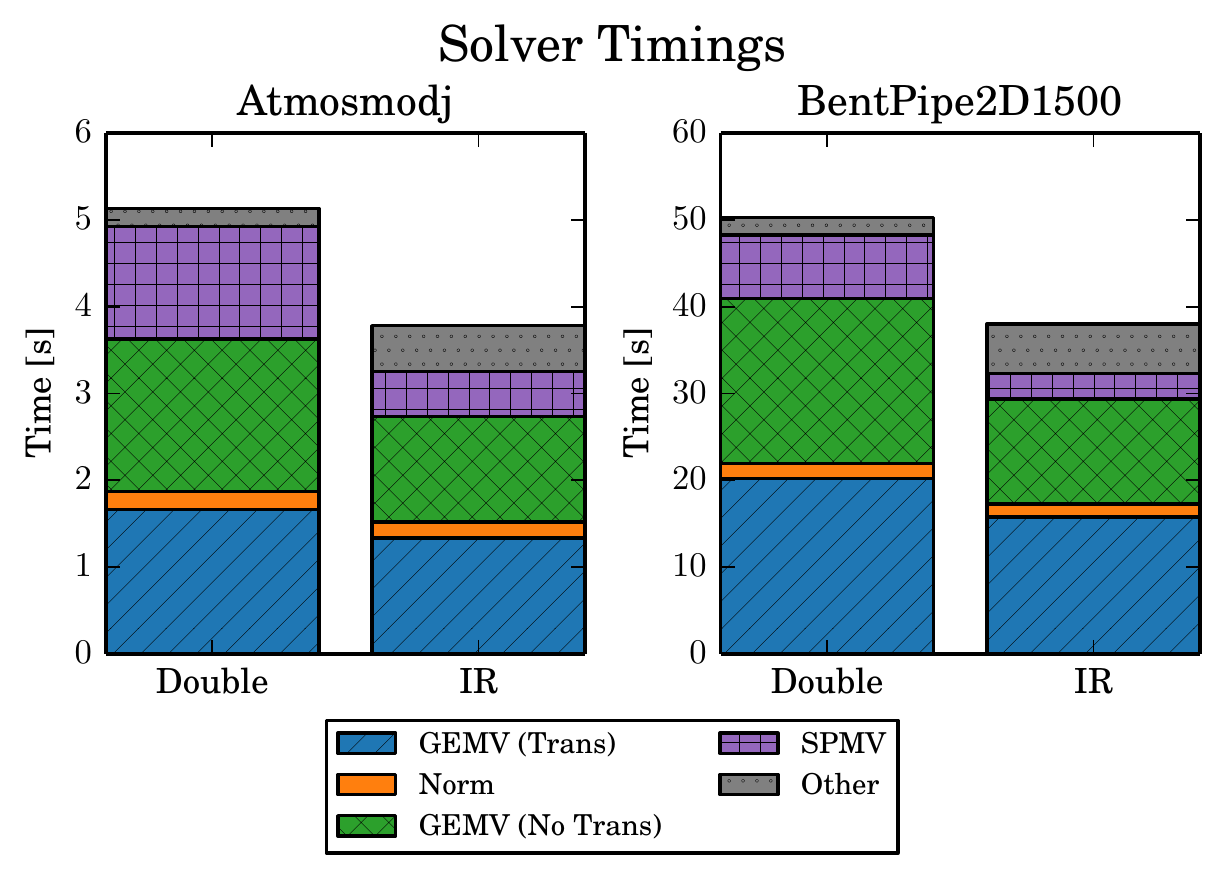}
  \caption{Solve times for GMRES$(50)$ double (left) and IR (right) for the matrix BentPipe2D1500. Each bar represents total solve time, split up to give a breakdown of time spent in different kernels. The ``Other" portion represents timing for small dense (non-GPU) operations and, for GMRES-IR, computing residuals in fp64.} 
  \label{fig:BentPipeTiming}
\end{figure}

Figure \ref{fig:BentPipeTiming} shows the solve times of the GMRES double and IR solvers, split over different kernels. The bar segment in Figure \ref{fig:BentPipeTiming} labeled ``other" indicates time solving the least squares problems and performing other non-GPU operations. For GMRES-IR, it also includes computation of the new iterative refinement residual in double precision. Solve times do not include time required to copy the matrix $A$ from fp64 to fp32 at the beginning of GMRES-IR. 
Table \ref{tab:BentPipeSpeedup} shows the speedup attained by different kernels for the double precision and IR solves with each matrix. By this measure, GMRES-IR gives $1.32\times$ to $1.36\times$ speedup over the solve time of GMRES double. The GEMV kernels give from $1.25$ to $1.57\times$ speedup, but the SpMV gives a spectacular $2.48\times$ speedup!
\begin{table}[htbp]
  \centering
  \caption{Speedup of different kernels from GMRES Double to GMERS-IR for the matrices Atmosmodj and BentPipe2D1500. (Note that this is speedup of the total time spent in each kernel in GMRES double vs GMRES-IR. This is not a per-call comparison.)}
    \resizebox{\columnwidth}{!}{%
    \begin{tabular}{lrr}
          & \multicolumn{1}{l}{Atmosmodj} & \multicolumn{1}{l}{BentPipe2D1500} \\
          \hline 
    \textbf{GEMV (Trans)} & 1.25  & 1.28 \\
    \textbf{Norm} & 1.13  & 1.15 \\
    \textbf{GEMV (No Trans)} & 1.45  & 1.57 \\
    \textbf{Total Orthogonalization } & 1.32  & 1.38 \\
    \textbf{SpMV} & 2.48  & 2.48 \\
    \textbf{Total time w/ refinement ops} & 1.36  & 1.32 \\
    \end{tabular}%
     } 
  \label{tab:BentPipeSpeedup}%
\end{table}%

\begin{figure}
  \centering
  \includegraphics[width=0.48\textwidth,keepaspectratio]{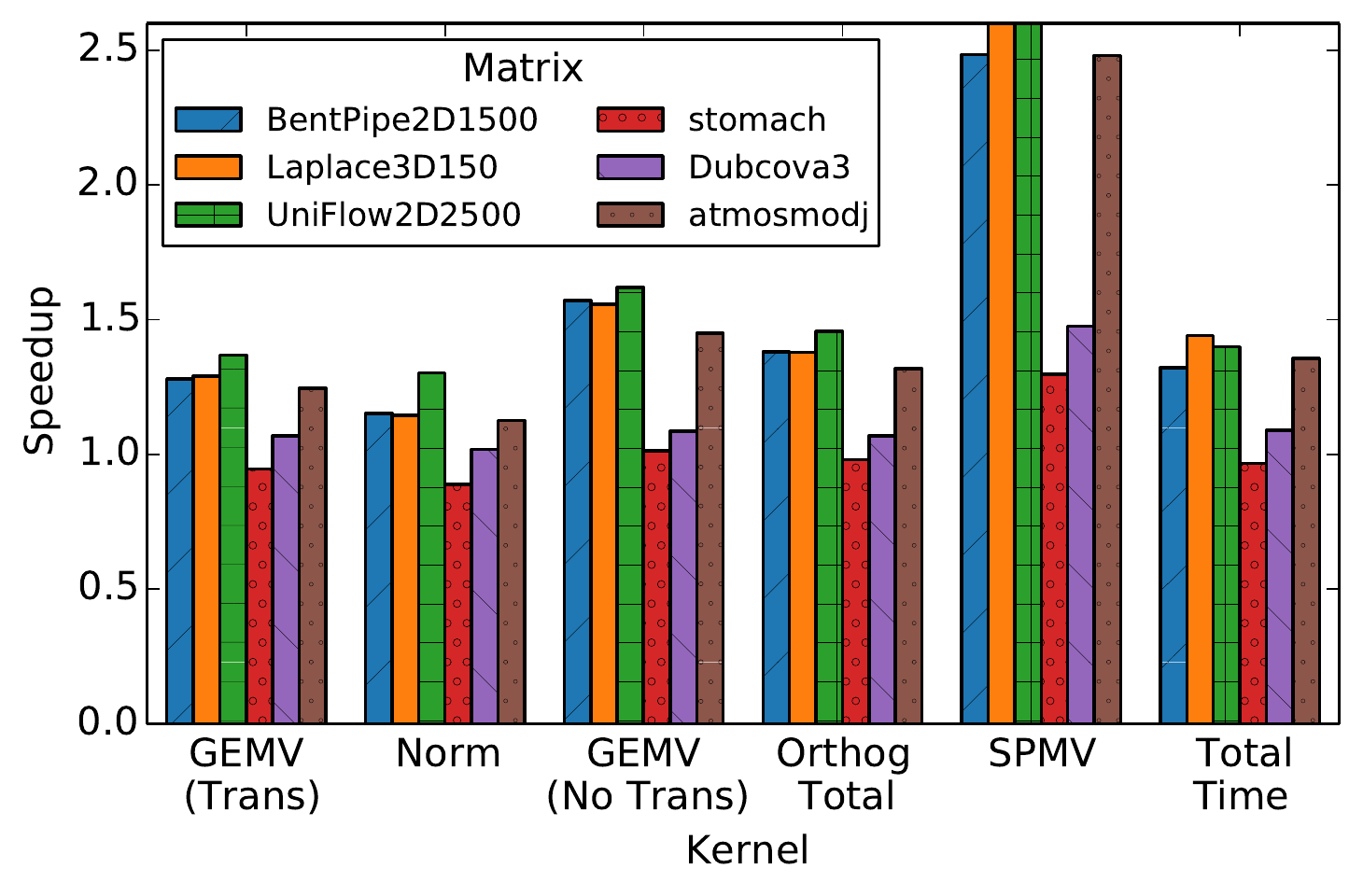}
  \caption{Speedups for different kernels going from GMRES double to GMRES-IR six different matrix problems. (Note that this is speedup of the total time spent in each kernel in GMRES double vs GMRES-IR. This is not a per-call comparison.)}
  \label{fig:KernelSpeedup3Mats}
\end{figure}

In Figure \ref{fig:KernelSpeedup3Mats}, we show kernel speedups for the previous problems and four additional matrices: the matrix UniFlow2D2500 from Section \ref{subsec:IRvsFD}, a 3D Laplacian with $nx=150$, and matrices stomach and Dubcova3 from SuiteSparse. 
(See Table \ref{tab:largeTestSet} for additional problem statistics.)
Note that the numbers in Table \ref{tab:BentPipeSpeedup} and the bars in Figure \ref{fig:KernelSpeedup3Mats} show the speedup of the entire time GMRES double spends in a kernel over the entire time GMRES-IR spends in the same kernel. Since GMRES-IR needs a few extra iterations (and kernel calls) beyond what GMRES double needs to converge, this is not a per-call time comparison. Even so, speedups for a per-call comparison are very similar to those presented in Figure \ref{fig:KernelSpeedup3Mats}.
It is interesting to note that the kernel speedups are relatively consistent across the three Galeri problems and atmosmodj. In particular, the SpMV kernel improves by $2.4$ to $2.6$ times in these four cases. This occurs due to near-perfect L2 cache reuse for the right-hand side vector with SpMV float, while there is a high L2 cache miss rate for SpMV double. We will discuss SpMV speedup further in Sections \ref{sec:SpMV} and \ref{sec:SpMVTestSet}.
The total solve times to convergence for these four problems improve by $24$ to $36\%$. For the Dubcova3 and stomach problems, however, speedups are less spectacular.  In fact, for stomach, the GEMV (Trans) and Norm kernels experience slowdown in the single precision solve within GMRES-IR.  
We are currently investigating the poor speedup of these kernels with teams from NVIDIA (for CuBlas) and Kokkos Kernels. We hope to find a solution in the near future. 


\subsection{GMRES-IR Convergence and Speedup Can Vary with Problem Right-Hand Side}
\label{sec:rhscomp}
Here we examine the (unpreconditioned) SuiteSparse problem parabolic\_fem, a convection-diffusion problem from computational fluid dynamics. We observe the problem convergence for GMRES double and GMRES-IR with four different right-hand sides: 1) a vector RHS\_Ones of all ones (as is used for the other problems in this paper), 2) the right-hand side vector given from SuiteSparse (RHS\_Given), 3) a vector RHS\_Unif with random entries that are uniformly distributed in the interval $(0,1)$, and 4) a vector with entries drawn from the standard normal distribution (RHS\_Norm). No additional scaling is performed. 

\begin{figure}
    \centering
    \includegraphics[width=0.45\textwidth,keepaspectratio]{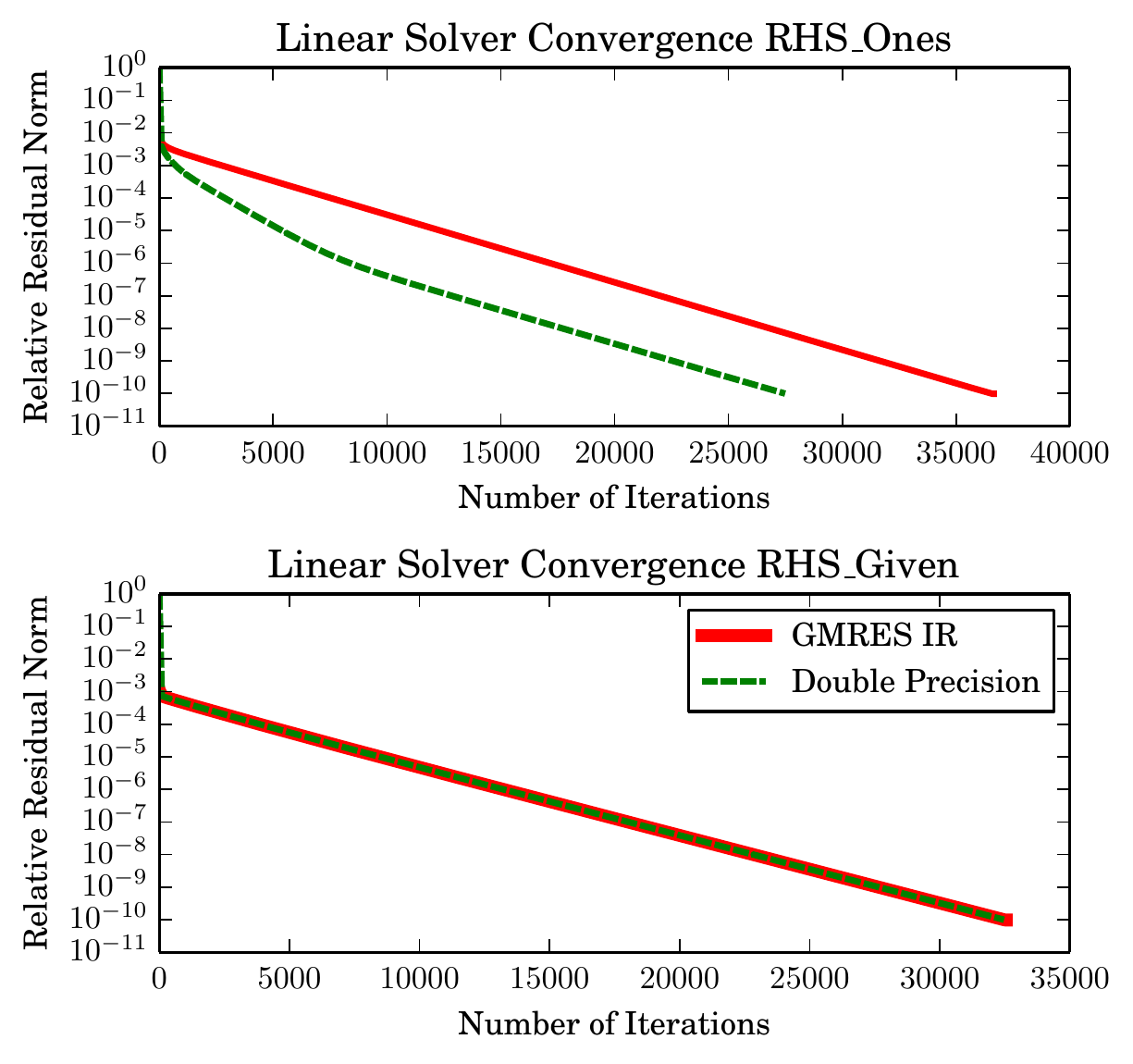} 
    \caption{GMRES double and GMRES-IR convergence curves for the matrix parabolic\_fem with two different right-hand sides. (Top) vector of all ones, (Bottom) given vector from SuiteSparse.  }
    \label{fig:parFemRhsConv}
\end{figure}

Unlike the previous two examples, for the RHS\_Ones vector, GMRES-IR convergence does not follow that of GMRES-Double. (See top of Figure \ref{fig:parFemRhsConv}.)
This likely occurs due to eigenvector components in the RHS\_Ones that fp32 GMRES cannot approximate well. Future investigation may be able to determine the specific relevant eigenvalues.
For the remaining three right-hand sides, however, convergence of GMRES-IR does follow that of GMRES Double.  (See the bottom of Figure \ref{fig:parFemRhsConv} for convergence with RHS\_Given.)  Table \ref{tab:parFemRhs} shows timings and iteration counts for all four configurations. The RHS\_Ones problem run time is slower with GMRES-IR due to the extra $9{,}107$ iterations it needs over the double precision solver. It is somewhat ironic that the problem which needs the fewest iterations to converge in double precision takes the most iterations with GMRES-IR. The remaining three problems achieve $1.25\times$ speedup or more with GMRES-IR.  
\begin{table}[htbp]
  \centering
  \caption{GMRES Double and GMRES-IR solve times and iteration counts for matrix parabolic\_fem with four different right-hand side vectors.}
   \resizebox{\columnwidth}{!}{%
    \begin{tabular}{l|rrrrc}
          & \multicolumn{2}{c}{\textbf{Double}} & \multicolumn{2}{c}{\textbf{IR}} &  \\
    \textbf{RHS Vec} & \multicolumn{1}{l}{\textbf{Time}} & \multicolumn{1}{l}{\textbf{Iters}} & \multicolumn{1}{l}{\textbf{Time}} & \multicolumn{1}{l}{\textbf{Iters}} & \multicolumn{1}{l}{\textbf{Speedup}} \\
    \hline
    RHS\_Ones & 42.39 & 27,493 & 44.63 & 36,600 & 0.95 \\
    RHS\_Given & 50.04 & 32,470 & 39.16 & 32,500 & 1.28 \\
    RHS\_Norm & 54.02 & 34,960 & 41.72 & 35,000 & 1.29 \\
    RHS\_Unif & 51.98 & 33,625 & 41.64 & 34,150 & 1.25 \\
    \end{tabular}%
    } 
  \label{tab:parFemRhs}%
\end{table}%

\subsection{Convergence and Kernel Speedup for Preconditioned GMRES vs GMRES-IR}
\label{subsec:precon_compar}
For this section, we consider a 2D Laplacian matrix over a stretched grid. 
It has a large condition number, so GMRES$(50)$ cannot converge without preconditioning. 
We apply a GMRES polynomial preconditioner \cite{LoePPTrilinos} of degree $40$, comparing three options: a) GMRES-double with double precision preconditioning, b) GMRES-double with single precision preconditioning, c) and GMRES-IR with single precision preconditioning. Here ``single precision preconditioning" indicates that the polynomial is both computed and applied in single precision. Note that applying an fp32 preconditioner to an fp64 vector results in a non-constant preconditioner.  This means that the convergence theory for GMRES no longer holds and that one should use FGMRES \cite{SaadFGMRES} for guaranteed convergence.  However, for this experiment with polynomial preconditioning, GMRES does not seem to suffer from the inexact preconditioner; thus, we do not include FGMRES here. 

\begin{figure}
  \centering
  \includegraphics[width=0.45\textwidth,keepaspectratio]{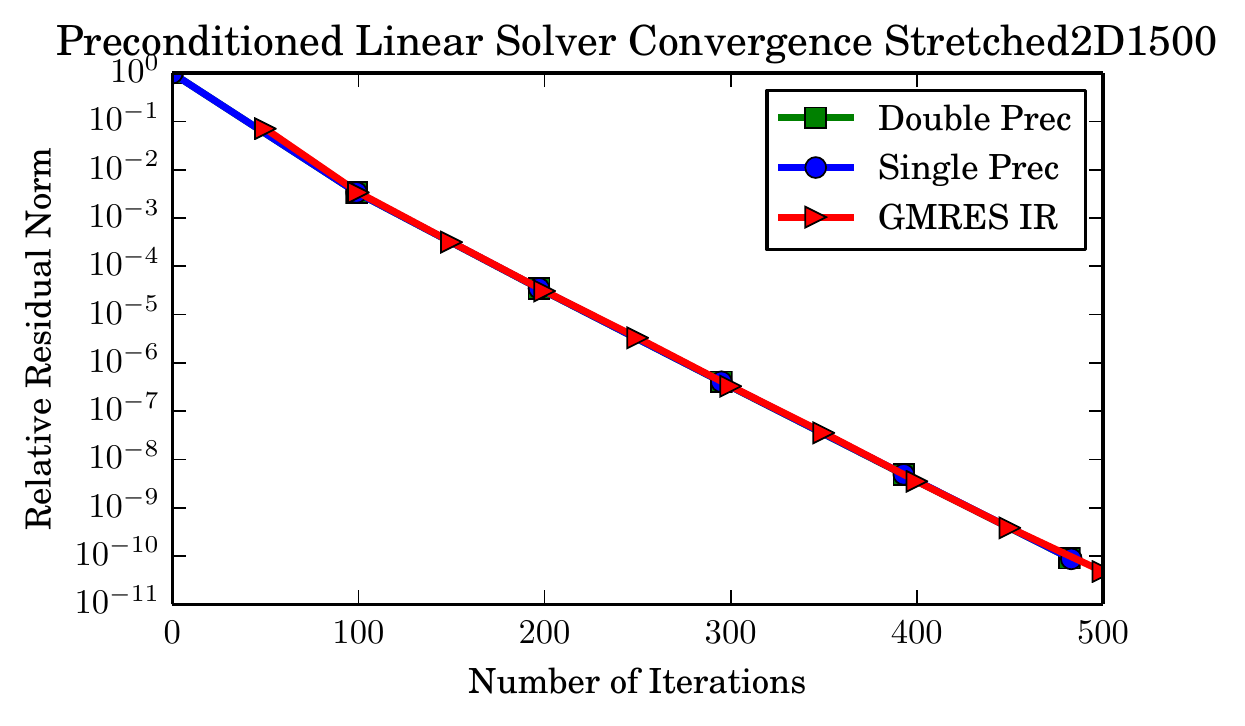}
  \caption{Convergence of the Stretched2D1500 problem with a degree $40$ polynomial preconditioner. Squares indicate fp64 preconditioning, circles fp32 preconditioning, and triangles GMRES-IR with fp32 preconditioning.}
  \label{fig:polyConv}
\end{figure}

\begin{figure}
  \centering
  \includegraphics[width=0.35\textwidth,keepaspectratio]{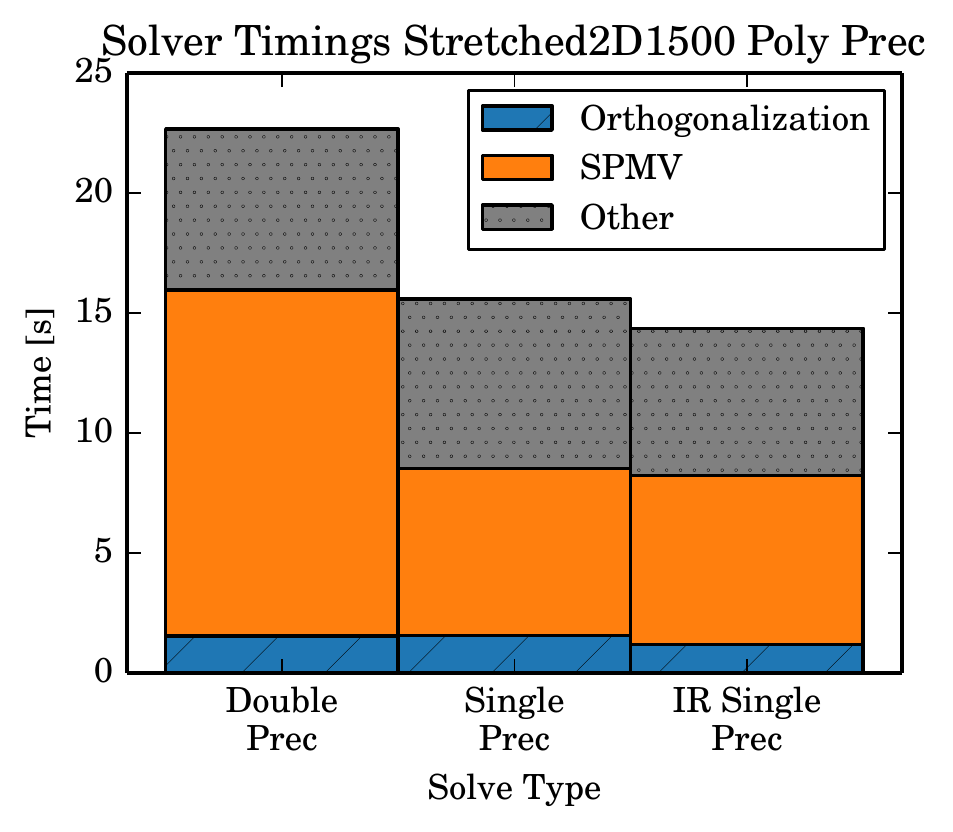}
  \caption{Solve times for polynomial preconditioned GMRES using polynomial degree $40$. The bar on the left shows solve time for fp64 GMRES, the bar in the middle shows fp64 GMRES with an fp32 polynomial, and the bar on the right gives timings for GMRES-IR with fp32 polynomial preconditioning.}
  \label{fig:PolyPrecKernelSpeedup}
\end{figure}

Figure \ref{fig:polyConv} demonstrates that, just as before, the problems with fp32 preconditioning converge very similarly to GMRES in fp64. 
Figure \ref{fig:PolyPrecKernelSpeedup} shows solve times for all three configurations. Times do not include creation of the polynomial preconditioner, which was $0.5$ seconds or less for all cases. 
Similar to Figure \ref{fig:BentPipeTiming}, the ``other" portion of each bar indicates time spent in dense matrix operations, vector additions for the polynomial, and computation of double-precision residuals in GMRES-IR. Since the SpMV constitutes the majority of kernel calls in the polynomial apply, the total SpMV time drops significantly in single precision as opposed to double. Time spent in ``other" operations, however, increases slightly due to the casting operations required to multiply an fp32 matrix polynomial with an fp64 vector. 
Ultimately, GMRES-IR gives $1.58\times$ speedup over GMRES double and $1.08\times$ speedup over simply preconditioning in float. Recall that GMRES-IR performs preconditioning and orthogonalization in low precision, giving it potential to have better performance than simply using a low precision preconditioner. Even when testing other polynomial degrees, the fp32 preconditioned GMRES gives reasonable speedup over the all-double precision GMRES, but run times are never faster than those of GMRES-IR.  

Unlike previous examples where solve time was dominated by orthogonalization, polynomial preconditioning shifts the cost toward the sparse matrix-vector product. Here, the SpMV gets about $2\times$ speedup going from fp64 to fp32. 
Note that in the previous example (Figure \ref{fig:BentPipeTiming}), the BentPipe SpMV kernel only comprises $15\%$ of the fp64 solve time, so the $2.5\times$ SpMV speedup only removes $4.4$ seconds from the original solve time of $50$ seconds. In this stretched Laplacian problem, the SpMV comprises $64\%$ of the total solve time for fp64 GMRES, so the improvement in SpMV time provides $32\%$ of the ultimate speedup in GMRES-IR. Polynomial preconditioning allows us to take advantage of the large speedup from applying the SpMV in lower precision. 


While this analysis has only covered polynomial preconditioning, we believe that the following concepts will also extend to many other preconditioners: a) The convergence of an fp64 GMRES solver does not necessarily suffer from using an fp32 preconditioner instead of an fp64 preconditioner; b) While using an fp32 preconditioner does not degrade the convergence of GMRES, it will typically improve solve time over using the same preconditioner in fp64;
and c) Preconditioning allows users to take advantage of kernels that have large speedup in lower precisions. 

For completeness, we discuss a case where single precision preconditioning does adversely affect GMRES convergence.
We test polynomial degrees that are multiples of $10$ up to $70$. 
For the previous Stretched2D1500 problem, both the GMRES-IR and the GMRES-FD (GMRES fp64$+ $fp32 polynomial preconditioner) solvers converge to 1e-10 for all polynomial degrees. However, a related matrix gives a different result. We precondition a $3$D Laplacian that has $200$ grid points in each direction, so it is significantly larger than the previous example and no longer has a stretched grid. When we apply the polynomial and all other operations in fp64, the GMRES solver always converges successfully for all polynomial degrees. Then we apply the polynomial preconditioner in fp32 and perform all other GMRES calculations in fp64. For the degree $10$ polynomial the solver converges, just as it does in all double precision. However, for higher degree polynomials, the implicit residual (that which results from applying Givens rotations to the matrix $H$ from the Arnoldi relation) diverges from the explicit residual (computed by forming $\hat{x}$ and calculating $\|b-A\hat{x}\|_2$). It is likely that this polynomial preconditioner becomes ill-conditioned far more quickly in single precision than in double. 

In the Belos solvers library, divergence of the implicit and explicit residuals is denoted as a ``loss of accuracy" of the solver. In essence, the solver gives a ``false positive" signal of convergence. With the degree $40$ polynomial, for instance, Belos stops after $32$ iterations because the implicit residual has reached $3.22\mathrm{e}{-}11$. However, the norm of the true residual is only $8.5\mathrm{e}{-}4$.  One can likely address this manually by re-running the Belos solver using the original solution vector as $x_0$, but Belos does not address loss of accuracy in this way automatically.  GMRES-IR, on the other hand, overcomes this barrier naturally due to its corrections at each restart in double precision.  With degree $40$ preconditioning, GMRES-IR converges to tolerance $1.8\mathrm{e}{-}12$ in $150$ iterations.  Unfortunately, the GMRES-IR solver is still slower than fp64 GMRES with fp64 preconditioning, which needs only $22$ iterations to converge.  In the future, we will investigate whether the flexible GMRES variant (FGMRES) can improve the solver accuracy when preconditioning in fp32.  

\subsection{Matrix Structure, Cache Reuse, and SpMV Performance}
\label{sec:SpMV}

The roughly $2.5\times$ speedup of the sparse matrix-vector product in the previous examples requires deeper explanation. Intuitively, one might expect that changing the working precision from fp64 to fp32 should give at most 1.5 to two times speedup since we are reducing the memory requirement by almost half. We assume the integer index type stays the same. If we halve the floating point data size and the index size stays the same, 
then one might expect at most $1.5 \times$ speedup. Below we explain how lower precision can improve cache reuse and give greater than $1.5$ or even $2$ times speedup. 

Note that the SpMV kernel called in all previous examples is an implementation native to Kokkos Kernels; we do not employ CuSparse for SpMV (though CuBlas may be called in other operations). The SpMV kernel is memory-bound, so the limiting factor in speed is how fast data can be moved through the memory hierarchy. 
Recall that storing a double requires $8$ bytes of memory and that both integers and floats require $4$ bytes of memory. Each of our matrices is stored in Compressed Sparse Row (CSR) format. With NVIDIA profiling tools, we observed that the L2 cache hit rate for the float SpMV was almost twice the hit rate for the double SpMV. This appears to be due to ``perfect caching" of the right-hand side vector $x$. Below we give a calculation to explain how this caching effect can account for $2.5\times$ speedup. 

Suppose that $A$ has $w$ nonzero elements per row and $n$ rows (so $nnz=w*n$) and that we are computing $Ax=y$. With the CSR matrix storage format, we have two vectors of length $nnz$ [one for the values of $A$ (denoted $A_{val}$) and another for the column indices (denoted $colId$)] and a vector of row pointers of length $n+1$. For this calculation, we ignore reads of the vector of row pointers and writes to $y$ since they account for only a small fraction of all memory traffic. To compute the dot product for each element in the solution vector $y$, we have to read one row of nonzeros $A$ and $w$ elements of $x$ which correspond to their locations. Thus the first dot product is 
\[\sum_{i=0}^{w-1} A_{val}[i] * x[colId[i]]. \]
Suppose now that in fp64, there is no cache reuse for the $x$ vector; we have to reread each element from device memory to cache every time we need it. Then to compute the SpMV, for each nonzero element in $A$ we read one double from $A$, one int (for $colId[i]$), and another double from $x$. In that case, the total number of reads to cache is 
\[n*w*[size(int) + 2* size(double)] = 20wn.\]
Next, we suppose that in fp32 there is ``perfect caching" of the $x$ vector. In other words, we only have to read $x$ from device memory once, and after an element is read into cache, it stays there until we do not need it any longer. In that case, the total number of reads to cache is 
\[n*w*[size(int) + size(float)] +n*size(float) = (8w+4)n.\]
Then the speedup going from double to float is
\[ \frac{20wn}{(8w+4)n} = \frac{5w}{2w+1}.\]
This ratio quickly approaches $2.5$ as $w$ grows. For the matrices in Section \ref{subsec:unprecon_compar}, the speedup as predicted by the model is slightly lower. Matrices UniFlow2D2500 and BentPipe2D1500 have $5$ nonzeros per row, so the expected speedup is $2.27\times$. With the Laplace3D150 matrix that has $7$ nonzeros per row, the expected speedup from this model is $2.33\times$. The observed speedup in all three cases was slightly higher than expected, probably due to additional improvements in $L1$ cache use. 

Additional experiments have confirmed this model: for nicely structured matrices, $2.5\times$ speedup can result from perfect cache reuse for $x$ in fp32, while some $x$ vector elements must be re-read into cache for fp64. 
Note that if $A$ has larger bandwidth, elements of $x$ may be accessed with less spatial locality, so $2.5\times$ speedup is not expected. The next section analyzes SpMV speedup for a large set of test problems. For an additional study of accelerating SpMV using mixed precision, see \cite{AhmadMPSpMV}.



\subsection{SpMV Speedup: A Large Test Set}
\label{sec:SpMVTestSet}

To further validate the model in Section \ref{sec:SpMV}, we test SpMV speedup for a large collection of SuiteSparse matrices as well as several PDE problems from the Trilinos Galeri package. The full list of matrices and SpMV run times can be found in Appendix A. The test set contains $67$ problems from the SuiteSparse matrix collection and $19$ problems from Galeri. For each test, first we run an un-timed warm-up loop containing the SpMV kernel so that setup overhead is excluded from the final timings. We then record the elapsed time from $1000$ calls to SpMV with a random vector.  Run times for the SpMV in both float and double precision are taken to be the minimum of three runs.  

\begin{figure}
    \centering
    \includegraphics[width=0.45\textwidth]{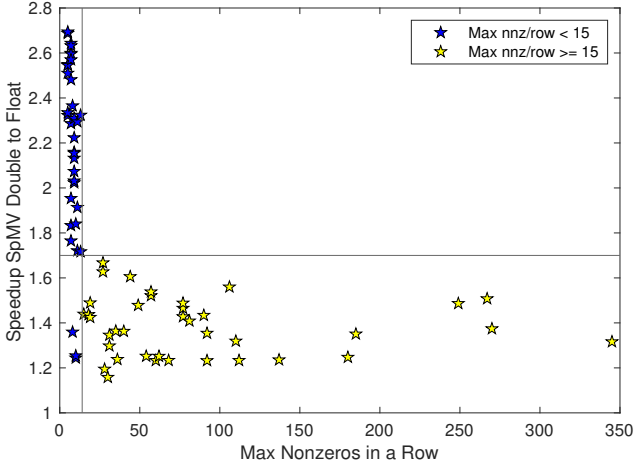}
    \caption{SpMV speedup from fp64 to fp32 for several matrices versus the maximum number of nonzeros in any row of the matrix. Dark stars indicate the matrices where the maximum number of nonzeros in a row was less than $15$.}
    \label{fig:SpMVSpeedup}
\end{figure}

For each matrix, Figure \ref{fig:SpMVSpeedup} plots the observed speedup of the SpMV versus the maximum number of nonzeros that the matrix has in any of its rows.  The maximum number of nonzeros in a row gives a rough estimate of the regularity of the matrix; that is, do all rows have a similar number of elements, or are there some rows that are relatively dense?  Figure \ref{fig:SpMVSpeedup} is divided into four quadrants with a vertical line at $x=15$ nonzeros and a horizontal line at $y=1.7\times$ speedup.  
Notice that all the matrices which attain $1.7\times$ speedup or greater are in the top left quadrant and have fewer than $15$ nonzeros in each row.  All matrices with $15$ or more nonzeros in the densest row (lower right quadrant) attained speedup less than $1.7\times$. We hypothesize that this could be due to the cache effects or reuse of the vector entries. Matrices with large maximum number of nonzeros in a row could results less cache reuse even for fp32. Note that there are $13$ matrices from Appendix A which are not plotted in Figure \ref{fig:SpMVSpeedup} because they have more than $350$ nonzeros in their densest rows.  These matrices also attained less than $1.7\times$ speedup. 

Observe that the bottom left quadrant shows three matrices with less than $15$ nonzeros in each row which did not attain $1.7\times$ speedup.  These three matrices are \textit{lung2, thermomech\_TC,} and \textit{thermomech\_TK}.  These matrices each had less than $720{,}000$ nonzeros in total, so that they were the smallest matrices in the test set.  We hypothesize that using lower precision SpMV provides little benefit for very small problems. One reason for this could be that even the double precision vector could potentially fit in the cache. A model that can also model cache effects could confirm this hypothesis in the future. 
While we did attempt to categorize SpMV speedup based upon matrix bandwidth as well, this measure proved to be less relevant than the maximum number of elements in the densest row.  Matrices with a small uniform number of nonzeros in each row are most likely to attain excellent speedup under this software model.  

Recall that the tested SpMV implementation is native to Kokkos Kernels \cite{RajamanickamKokkosKernels}. Results are likely to differ using CuSparse or other sparse math libraries as the performance is implementation dependent. The Kokkos Kernels SpMV implementation does not currently attempt to use sophisticated optimizations for matrices with dense rows; with algorithm advancements, the speedup for matrices with dense rows can likely be improved.


\subsection{Choosing a Restart Size for GMRES-IR}
\label{subsec:restart}
Here we demonstrate an interesting case where choosing a small restart size for GMRES gives improved performance (in double and mixed precisions) over a large subspace size.
The authors of \cite{LindquistGMRES} devote many experiments to determining the best restart strategy for GMRES-IR. Their strategy is to pick the restart size that allows the inner low precision GMRES to converge as far as possible before restarting. This means picking the largest subspace possible before convergence in the inner solver stalls. Here we demonstrate a further example with matrix BentPipe2D1500 where the large matrix size causes orthogonalization costs to dominate the solve time. Thus, a smaller restart size is more beneficial for this problem. 

We test a variety of restart sizes; Table \ref{tab:ManySubsBentPipe} gives the solve times and iteration counts. In each case, GMRES-IR still gives speedup of $1.20\times$ to $1.40\times$ over GMRES double. 
\begin{table}[htbp]
 \centering
 \caption{BentPipe2D1500 Convergence for Many Restart Sizes}
  \resizebox{\columnwidth}{!}{%
  \begin{tabular}{llrrrr}
  \multicolumn{1}{p{4.085em}}{\textbf{Subsp }} & \multicolumn{2}{c}{\textbf{GMRES Double}} & \multicolumn{2}{c}{\textbf{GMRES-IR}} & \\
  \multicolumn{1}{p{4.085em}}{\textbf{Size}} & Iters & Solve Time & Iters & Solve Time & \multicolumn{1}{l}{Speedup} \\
  \hline
  25  & 13795 & 38.63 & 13925 & 31.74 & 1.22 \\
  50  & 12967 & 50.26 & 13150 & 38.03 & 1.32 \\
  100  & 12009 & 74.24 & 12100 & 51.88 & 1.43 \\
  150  & 11250 & 95.82 & 12450 & 72.01 & 1.33 \\
  200  & 10867 & 117.80 & 12400 & 90.77 & 1.30 \\
  300  & 10491 & 164.60 & 12600 & 133.60 & 1.23 \\
  400  & 10274 & 209.80 & 12400 & 174.10 & 1.21 \\
  
  \end{tabular}%
  } 
 \label{tab:ManySubsBentPipe}%
 
\end{table}%
Although the iteration count for the fp64 solver decreases as the subspace gets larger, the solve time increases. The large subspace size causes orthogonalization costs to increase and dominate the solve time more and more. Figure \ref{fig:BentPipeTiming} demonstrates the proportion of total orthogonalization costs (GEMV Trans $+$ Norm $+$ GEMV no Trans) for restart size $50$. In that figure, orthogonalization consumes $83\%$ of solve time for the fp64 solver and $80\%$ of solve time for GMRES-IR. As the restart size increases, the proportion of solve time for SpMVs and non-orthogonalization operations gets squeezed out. With a restart size of $400$, orthogonalization takes $97\%$ of solve time for both the double precision and IR solvers. 

The smallest restart size of $25$ also gives the best solve time for GMRES-IR. Like the double precision solver, GMRES-IR benefits from reduced orthogonalization costs with the small restart length. Observe that, contrary to the strategy in \cite{LindquistGMRES}, size $25$ gives us the fastest solve time even though the single precision inner solver convergence is not near stalling. Even for the largest restart size of $400$, the residuals of the inner solver do not appear to have stalled; they are all on the order of $0.1$. Typically an fp32 solver can converge to near $10^{-5}$ without any fp64 refinement. 

As the inner fp32 GMRES solver is restarted (and refined) less frequently, the gap between the iteration count needed for GMRES-IR and GMRES double convergence widens. Note that, while unusual, accumulated rounding errors can occasionally help GMRES-IR to need fewer iterations to converge than GMRES double. This phenomenon did not happen in the presented median-time runs, but it was occasionally present in other runs of the experiment. Nevertheless, the GMRES-IR solver consistently gives performance improvement over the GMRES double solver. 

\begin{figure}
  \centering
  \includegraphics[width=0.5\textwidth,keepaspectratio]{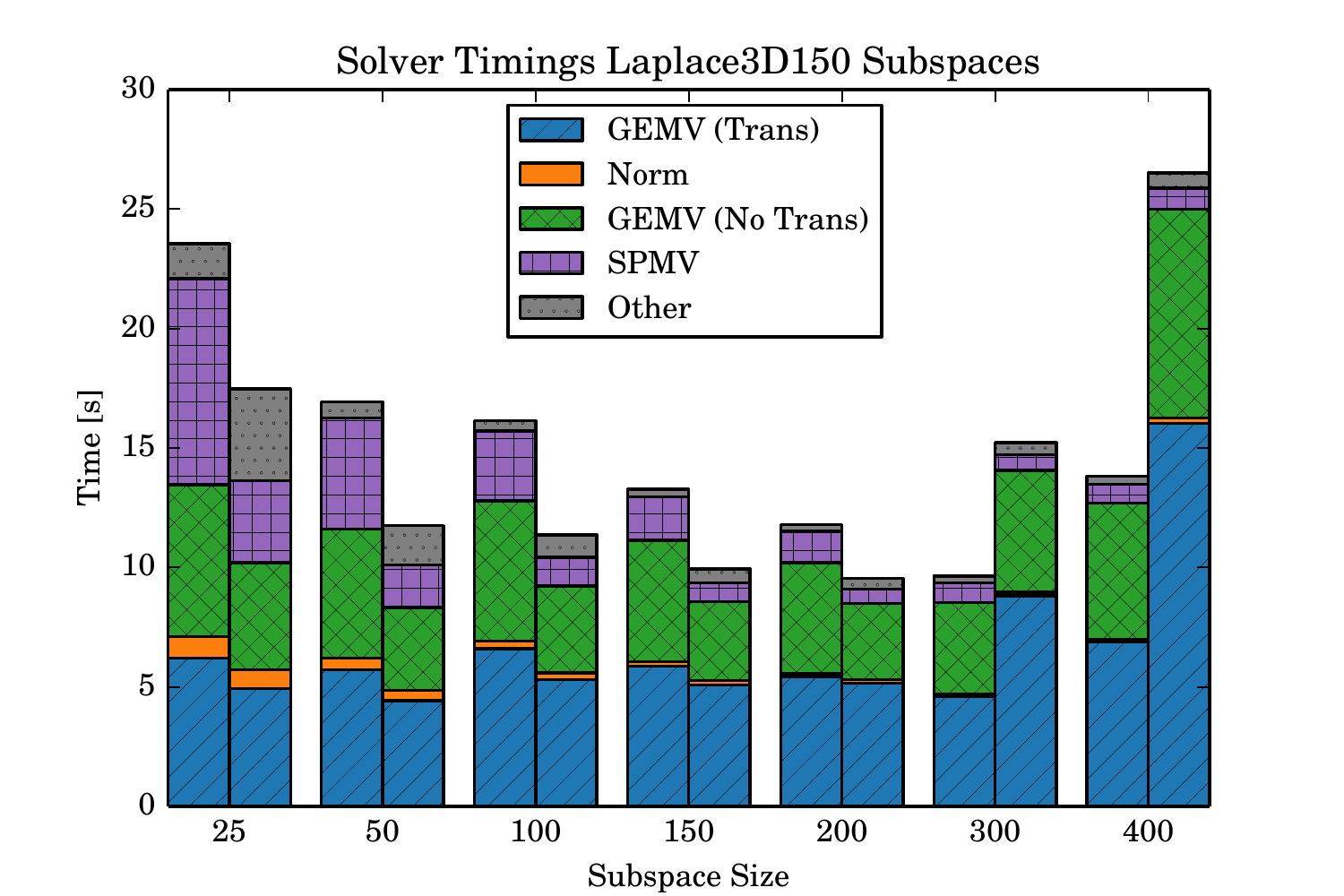}
  \caption{Total solve times for different GMRES restart lengths for the matrix Laplace3D150. For each restart size, the left bar indicates solve time for GMRES (double) and the right bar gives solve time for GMRES-IR. }
  \label{fig:LaplSubspMulti}
\end{figure}

Next, we show an example using a 3D Laplacian where GMRES-IR does not give speedups at large subspace sizes. 
Results are in Figure \ref{fig:LaplSubspMulti}, where bars indicating solve time are broken down into the times for particular kernels. 
For restart sizes up to $200$, the GMRES-IR solver gives $19\%$ to $31\%$ improvement in solve time over GMRES double. 
However, with larger subspaces, the iterative refinement solver needs so many additional iterations over GMRES double that we do not see any speedup. With size $300$, GMRES double needs $433$ iterations compared to $900$ iterations for GMRES-IR. For subspace size $400$, GMRES-IR needs almost three times as many iterations as GMRES double. In the experiments for both of these large subspace sizes, we see strong evidence of stalled convergence in the single precision solver; several residuals are on the order of $10^{-7}$. Slowdown comes with GMRES-IR because the double precision residual is updated so infrequently; the inner solver is taking extra iterations without making progress towards the solution. 

Ultimately, the fastest solve time is with GMRES-IR and a subspace size of $200$ (though the timing of GMRES$(300)$ in double was faster on some runs). 
It should be noted that for larger versions of this PDE matrix, attempting to use a subspace size of $300$ results in an out-of-memory error on the GPU. Thus, GMRES-IR likely gives the most practical gains in terms of solve time for large problems.  

\begin{table*}[htbp]
  \centering
  \caption{Timings and iteration counts for GMRES double and GMRES-IR for a variety of Suitesparse and Galeri matrices. In the ``symm" column, 'y' indicates a symmetric matrix and 'spd' indicates a symmetric positive definite matrix. In the ``Prec" column, ``J $k$" indicates block Jacobi preconditioning with block size $k$, and ``p $k$" indicates polynomial preconditioning of degree $k$.}
    \begin{tabular}{rlrrll|rr|rr|c}
          &       &       &       &       &       & \multicolumn{2}{c}{\textbf{Double}} & \multicolumn{2}{c}{\textbf{IR}} &  \\
    \multicolumn{1}{l}{\textbf{UF id}} & \textbf{Matrix Name} & \multicolumn{1}{l}{\textbf{N}} & \multicolumn{1}{l}{\textbf{NNZ}} & \textbf{symm} & \textbf{prec} & \multicolumn{1}{l}{\textbf{Time}} & \multicolumn{1}{l}{\textbf{Iters}} & \multicolumn{1}{l}{\textbf{Time}} & \multicolumn{1}{l}{\textbf{Iters}} & \multicolumn{1}{l}{\textbf{Speedup}} \\
    \hline
    2266  & atmosmodj & 1,270,432 & 8,814,880 & n     &       & 5.12  & 1740  & 3.78  & 1750  & \cellcolor[rgb]{ .604,  .816,  .529} 1.35 \\
    2267  & atmosmodl & 1,489,752 & 10,319,760 & n     &       & 1.61  & 446   & 1.23  & 450   & \cellcolor[rgb]{ .647,  .827,  .537} 1.31 \\
    1858  & crashbasis & 160,000 & 1,750,416 & n     &       & 0.55  & 431   & 0.52  & 450   & \cellcolor[rgb]{ .875,  .898,  .588} 1.07 \\
    1849  & Dubcova3 & 146,698 & 3,636,643 & spd   &       & 1.15  & 1131  & 1.05  & 1150  & \cellcolor[rgb]{ .851,  .89,  .58} 1.10 \\
    1852  & FEM\_3D\_thermal2 & 147,900 & 3,489,300 & n     &       & 0.84  & 775   & 0.80  & 800   & \cellcolor[rgb]{ .898,  .906,  .592} 1.05 \\
    1853  & parabolic\_fem & 525,825 & 3,674,625 & spd   &       & 42.39 & 27493 & 44.63 & 36600 & \cellcolor[rgb]{ .992,  .937,  .612} 0.95 \\
    1367  & SiO2  & 155,331 & 11,283,503 & y     &       & 18.23 & 17385 & 16.86 & 17600 & \cellcolor[rgb]{ .867,  .898,  .584} 1.08 \\
    895   & stomach & 213,360 & 3,021,648 & n     &       & 0.51  & 359   & 0.52  & 400   & \cellcolor[rgb]{ .965,  .929,  .608} 0.98 \\
    2259  & thermomech\_dM & 204,316 & 1,423,116 & y     &       & 0.27  & 88    & 0.27  & 100   & \cellcolor[rgb]{ .945,  .922,  .604} 1.00 \\
    894   & lung2 & 109,460 & 492,564 & n     & j 1   & 0.46  & 206   & 0.49  & 250   & \cellcolor[rgb]{ 1,  .937,  .612} 0.94 \\
    1266  & hood  & 220,542 & 9,895,422 & spd   & j 42  & 13.98 & 5762  & 9.04  & 5000  & \cellcolor[rgb]{ .42,  .757,  .49} 1.55 \\
    805   & cfd2  & 123,440 & 3,085,406 & spd   & p 25  & 6.05  & 1092  & 4.55  & 1100  & \cellcolor[rgb]{ .627,  .82,  .533} 1.33 \\
    1431  & filter3D & 106,437 & 2,707,179 & y     & p 25  & 25.24 & 4449  & 18.12 & 4450  & \cellcolor[rgb]{ .565,  .804,  .522} 1.39 \\
    2649  & Transport & 1,602,111 & 23,487,281 & n     & p 25  & 8.35  & 339   & 8.73  & 450   & \cellcolor[rgb]{ .984,  .933,  .612} 0.96 \\
          & BentPipe2D1500 & 2,250,000 & 11,244,000 & n     &       & 50.26 & 12967 & 38.03 & 13150 & \cellcolor[rgb]{ .635,  .824,  .537} 1.32 \\
          & Laplace3D150 & 3,375,000 & 23,490,000 & spd   &       & 16.93 & 2387  & 11.75 & 2400  & \cellcolor[rgb]{ .522,  .788,  .514} 1.44 \\
          & UniFlow2D2500 & 6,250,000 & 31,240,000 & n     &       & 29.62 & 2905  & 21.17 & 3000  & \cellcolor[rgb]{ .561,  .8,  .522} 1.40 \\
          & Stretched2D1500  & 2,250,000 & 20,232,004 & spd   & p 40  & 22.66 & 482   & 14.37 & 500   & \cellcolor[rgb]{ .388,  .745,  .482} 1.58 \\
          \hline
       \end{tabular}%
       
\label{tab:largeTestSet}%
\end{table*}%

\subsection{Testing GMRES-IR on Matrices from SuiteSparse}
\label{subsec:suitesparse}

Finally, we validate the prior analysis with additional examples. We test several matrices from the SuiteSparse matrix collection \cite{DavisSparseCollect} with GMRES double and GMRES-IR. Results are in Table \ref{tab:largeTestSet}. The first nine matrices do not have preconditioning. The next two matrices are reordered with a reverse Cuthill-McKee ordering before applying block Jacobi preconditioners with block sizes of $1$ and $42$, respectively. The next three matrices in the table use polynomial preconditioners of degree $25$. At the end of the table, we repeat the earlier results of Section \ref{sec:experiments} for completeness. 

Based on this test set, results for speedup from GMRES-IR are mixed. For this test set, it seems that GMRES-IR gave the best speedup for structured PDE problems and for problems with polynomial preconditioning. Other problems, such as stomach and Dubcova3, suffer from the poorly performing orthogonalization kernels first presented in Figure \ref{fig:KernelSpeedup3Mats}.  Recall that the original BentPipe2D1500 and atmosmodj problems from Section \ref{subsec:unprecon_compar} gained most benefit with GMRES-IR because of the speedup in orthogonalization time.  To make GMRES-IR beneficial for all problems, it will be important to make sure that GEMV functions take advantage of fp32 precision for performance gains with all sizes of matrices.  

In problems where we do see speedup from GMRES-IR, the values vary from $1.05\times$ to $1.58\times$. Typically GMRES$(50)$-IR needs a few more iterations to converge than GMRES$(50)$ double, but the \textit{hood} matrix is a counterexample. For the \textit{hood} matrix, roundoff errors allow GMRES-IR to converge with $762$ fewer iterations than GMRES double, giving us a higher speedup than expected from simply switching to a lower working precision.

In all, 13 of 18, or 72\% of matrices experienced at least some solve time improvement from using GMRES-IR. Slowdowns, where they did occur, were minimal. This suggests that GMRES-IR can be a sufficient alternative to GMRES in fp64 if needed for working on a computer with low-precision hardware.  




\section{Conclusions and Future Work}
In this work, we evaluated two different approaches for mixed precision GMRES. We found GMRES-IR to be the best choice. GMRES-IR is a flexible algorithm for incorporating lower precision calculations into GMRES while maintaining double precision accuracy of the final solution. We demonstrate that using low precision arithmetic in GMRES-IR typically results similar convergence to the double precision solver. We observed speedups of up to $1.5\times$ using GMRES-IR with polynomial preconditioning. We analyzed the speedup at the individual kernel level, giving a model for speedup of the sparse matrix-vector product and recommended guiding principles for selecting solver restart length. 

Ongoing work is examining GMRES-IR with Tpetra linear algebra, which can support distributed memory parallel computing.  We are analyzing the mixed precision solver's capabilities and scalability on multiple GPUs (multi-node) and with a larger variety of preconditioners.
Another preliminary effort studies incorporating half precision into the GMRES-IR solve.  Initial attempts at running the inner GMRES in half precision have quickly resulted in numerical overflow issues with kernel reductions in the $2$-norm and triangular solve kernels.  This can likely (in part) be remedied by using a scaling strategy such as the one presented in \cite{HighamSqueezeHalf}.  
Future research will also evaluate GMRES-IR on non-NVIDIA architectures, such as the AMD MI100 GPU.  It is important to determine the value of mixed precision solvers on architectures relevant to upcoming and future HPC systems. 
Software development efforts are in progress to make GMRES-IR available for Trilinos users in the Belos solvers package. We believe this new solver implementation could replace standard (all double) GMRES in many applications. 

\section*{Acknowledgment}
Thanks to Christian Trott and Luc Berger-Vergiat for helping develop the model in Section V-D. We also thank the referees for their may useful suggestions. 
This research was supported by the Exascale Computing Project (17-SC-20-SC), a collaborative effort of the U.S. Department of Energy Office of Science and the National Nuclear Security Administration.


\bibliographystyle{IEEEtran}
\bibliography{IEEEabrv,MultiPrecisionNoURL.bib}

\appendix 
\clearpage
\onecolumn
\section{Sparse Matrices Tested}
%
\begin{table*}[h]
  \centering
  \smallfont
  \caption{Matrices tested for SpMV speedup from double to float.  For SuiteSparse matrices the matrix ID (`SSID') is listed.  Matrices with blank `SSID' are PDE problems from the Trilinos Galeri package.  The `Double' and `Float' columns list the run time of $1000$ iterations of SpMV in fp64 and fp32, respectively.  `Speedup' indicates the time for fp64 SpMV over the time for fp32 SpMV.  }
  \resizebox{\textwidth}{!}{%
    \begin{tabular}{lrrrr}
    \textbf{Matrix} & \multicolumn{1}{l}{\textbf{SSID}} & \multicolumn{1}{l}{\textbf{Double}} & \multicolumn{1}{l}{\textbf{Float}} & \multicolumn{1}{l}{\textbf{Speedup}} \\
    \hline
    2cubes\_sphere & 1919  & 0.066 & 0.051 & \cellcolor[rgb]{ .918,  .914,  .596} 1.30 \\
    af\_0\_k101 & 1580  & 0.341 & 0.250 & \cellcolor[rgb]{ .894,  .906,  .592} 1.36 \\
    af\_shell1 & 940   & 0.342 & 0.251 & \cellcolor[rgb]{ .894,  .906,  .592} 1.36 \\
    analytics & 2851  & 7.457 & 6.135 & \cellcolor[rgb]{ .949,  .922,  .604} 1.22 \\
    apache2 & 1423  & 0.400 & 0.169 & \cellcolor[rgb]{ .514,  .784,  .51} 2.37 \\
    atmosmodj & 2266  & 0.729 & 0.284 & \cellcolor[rgb]{ .435,  .761,  .494} 2.57 \\
    atmosmodl & 2267  & 0.839 & 0.319 & \cellcolor[rgb]{ .412,  .753,  .49} 2.63 \\
    audikw\_1 & 1252  & 1.357 & 1.031 & \cellcolor[rgb]{ .914,  .91,  .596} 1.32 \\
    Baumann & 1855  & 0.086 & 0.047 & \cellcolor[rgb]{ .718,  .851,  .553} 1.83 \\
    BenElechi1 & 1850  & 0.245 & 0.195 & \cellcolor[rgb]{ .937,  .918,  .6} 1.25 \\
    bmw7st\_1 & 1253  & 0.133 & 0.107 & \cellcolor[rgb]{ .941,  .922,  .6} 1.24 \\
    bundle\_adj & 2664  & 1.259 & 1.053 & \cellcolor[rgb]{ .957,  .925,  .604} 1.20 \\
    c-big & 1579  & 3.028 & 2.125 & \cellcolor[rgb]{ .871,  .898,  .584} 1.42 \\
    cfd2  & 805   & 0.086 & 0.074 & \cellcolor[rgb]{ .973,  .929,  .608} 1.16 \\
    cop20k\_A & 2378  & 0.107 & 0.076 & \cellcolor[rgb]{ .878,  .898,  .588} 1.41 \\
    crashbasis & 1858  & 0.118 & 0.061 & \cellcolor[rgb]{ .686,  .839,  .545} 1.91 \\
    CurlCurl\_1 & 2570  & 0.123 & 0.072 & \cellcolor[rgb]{ .761,  .863,  .561} 1.72 \\
    dc1   & 1320  & 18.460 & 13.770 & \cellcolor[rgb]{ .902,  .91,  .592} 1.34 \\
    d\_pretok & 1231  & 0.096 & 0.056 & \cellcolor[rgb]{ .757,  .863,  .561} 1.72 \\
    Dubcova3 & 1849  & 0.152 & 0.103 & \cellcolor[rgb]{ .851,  .89,  .58} 1.48 \\
    ecology2 & 1883  & 0.264 & 0.113 & \cellcolor[rgb]{ .525,  .788,  .514} 2.33 \\
    Emilia\_923 & 2542  & 0.812 & 0.534 & \cellcolor[rgb]{ .835,  .886,  .576} 1.52 \\
    F1    & 1411  & 0.535 & 0.385 & \cellcolor[rgb]{ .886,  .902,  .588} 1.39 \\
    Fault\_639 & 2543  & 0.578 & 0.384 & \cellcolor[rgb]{ .839,  .886,  .58} 1.51 \\
    FEM\_3D\_thermal2 & 1852  & 0.138 & 0.083 & \cellcolor[rgb]{ .78,  .871,  .565} 1.67 \\
    filter3D & 1431  & 0.086 & 0.070 & \cellcolor[rgb]{ .945,  .922,  .6} 1.23 \\
    Ga10As10H30 & 1350  & 0.151 & 0.129 & \cellcolor[rgb]{ .969,  .929,  .608} 1.17 \\
    Geo\_1438 & 2545  & 1.248 & 0.812 & \cellcolor[rgb]{ .827,  .886,  .576} 1.54 \\
    Goodwin\_095 & 2825  & 0.096 & 0.077 & \cellcolor[rgb]{ .937,  .918,  .6} 1.25 \\
    gsm\_106857 & 2329  & 0.671 & 0.430 & \cellcolor[rgb]{ .82,  .882,  .576} 1.56 \\
    hood  & 1266  & 0.250 & 0.175 & \cellcolor[rgb]{ .871,  .898,  .584} 1.43 \\
    hvdc2 & 1875  & 0.071 & 0.058 & \cellcolor[rgb]{ .945,  .922,  .6} 1.23 \\
    imagesensor & 2836  & 0.077 & 0.054 & \cellcolor[rgb]{ .871,  .898,  .584} 1.42 \\
    kkt\_power & 1876  & 0.600 & 0.419 & \cellcolor[rgb]{ .867,  .898,  .584} 1.43 \\
    ldoor & 1268  & 0.978 & 0.657 & \cellcolor[rgb]{ .847,  .89,  .58} 1.49 \\
    Lin   & 1213  & 0.167 & 0.085 & \cellcolor[rgb]{ .671,  .835,  .545} 1.95 \\
    lung2 & 894   & 0.048 & 0.035 & \cellcolor[rgb]{ .894,  .906,  .592} 1.36 \\
    mac\_econ\_fwd500 & 2376  & 0.118 & 0.073 & \cellcolor[rgb]{ .804,  .878,  .573} 1.61 \\
    marine1 & 2849  & 0.159 & 0.110 & \cellcolor[rgb]{ .867,  .898,  .584} 1.44 \\
    msdoor & 1644  & 0.440 & 0.301 & \cellcolor[rgb]{ .855,  .894,  .584} 1.46 \\
    nlpkkt80 & 1901  & 0.525 & 0.439 & \cellcolor[rgb]{ .957,  .925,  .604} 1.19 \\
    offshore & 2283  & 0.135 & 0.100 & \cellcolor[rgb]{ .902,  .906,  .592} 1.35 \\
    parabolic\_fem & 1853  & 0.361 & 0.158 & \cellcolor[rgb]{ .545,  .796,  .518} 2.29 \\

    \end{tabular}%

    \begin{tabular}{|lrrrr}
    \textbf{Matrix} & \multicolumn{1}{l}{\textbf{SSID}} & \multicolumn{1}{l}{\textbf{Double}} & \multicolumn{1}{l}{\textbf{Float}} & \multicolumn{1}{l}{\textbf{Speedup}} \\
    \hline 
    PFlow\_742 & 2661  & 0.672 & 0.544 & \cellcolor[rgb]{ .941,  .922,  .6} 1.24 \\
    power197k & 2816  & 0.220 & 0.178 & \cellcolor[rgb]{ .941,  .922,  .6} 1.24 \\
    power9 & 2838  & 0.346 & 0.258 & \cellcolor[rgb]{ .902,  .91,  .592} 1.34 \\
    PR02R & 2336  & 0.172 & 0.140 & \cellcolor[rgb]{ .945,  .922,  .6} 1.23 \\
    pwtk  & 369   & 0.222 & 0.178 & \cellcolor[rgb]{ .937,  .918,  .6} 1.25 \\
    Raj1  & 1863  & 6.226 & 4.273 & \cellcolor[rgb]{ .859,  .894,  .584} 1.46 \\
    rajat21 & 1370  & 19.100 & 13.370 & \cellcolor[rgb]{ .871,  .898,  .584} 1.43 \\
    Serena & 2541  & 1.301 & 0.875 & \cellcolor[rgb]{ .847,  .89,  .58} 1.49 \\
    ship\_003 & 1278  & 0.161 & 0.119 & \cellcolor[rgb]{ .898,  .906,  .592} 1.35 \\
    shipsec1 & 1279  & 0.154 & 0.125 & \cellcolor[rgb]{ .945,  .922,  .6} 1.23 \\
    SiO2  & 1367  & 0.308 & 0.279 & \cellcolor[rgb]{ .992,  .937,  .612} 1.10 \\
    stomach & 895   & 0.098 & 0.066 & \cellcolor[rgb]{ .847,  .89,  .58} 1.49 \\
    TEM152078 & 2812  & 0.154 & 0.117 & \cellcolor[rgb]{ .91,  .91,  .596} 1.32 \\
    thermal2 & 1403  & 0.627 & 0.273 & \cellcolor[rgb]{ .541,  .796,  .518} 2.29 \\
    thermomech\_dM & 2259  & 0.125 & 0.068 & \cellcolor[rgb]{ .714,  .847,  .553} 1.84 \\
    thermomech\_TC & 2257  & 0.061 & 0.049 & \cellcolor[rgb]{ .941,  .918,  .6} 1.24 \\
    thermomech\_TK & 2258  & 0.062 & 0.049 & \cellcolor[rgb]{ .937,  .918,  .6} 1.25 \\
    tmt\_sym & 1899  & 0.453 & 0.196 & \cellcolor[rgb]{ .537,  .792,  .514} 2.31 \\
    torso1 & 896   & 0.305 & 0.283 & \cellcolor[rgb]{ 1,  .937,  .612} 1.08 \\
    transient & 2275  & 9.947 & 7.422 & \cellcolor[rgb]{ .902,  .91,  .592} 1.34 \\
    Transport & 2649  & 0.507 & 0.352 & \cellcolor[rgb]{ .867,  .898,  .584} 1.44 \\
    twotone & 286   & 0.095 & 0.071 & \cellcolor[rgb]{ .898,  .906,  .592} 1.35 \\
    x104  & 1290  & 0.194 & 0.141 & \cellcolor[rgb]{ .89,  .906,  .588} 1.37 \\
    xenon2 & 802   & 0.146 & 0.090 & \cellcolor[rgb]{ .796,  .875,  .569} 1.63 \\
    BentPipe2D1500 &       & 0.556 & 0.218 & \cellcolor[rgb]{ .443,  .765,  .494} 2.55 \\
    Biharmonic2D1000 &       & 0.805 & 0.347 & \cellcolor[rgb]{ .529,  .792,  .514} 2.32 \\
    Laplace2D1000 &       & 0.265 & 0.114 & \cellcolor[rgb]{ .529,  .792,  .514} 2.32 \\
    Laplace2D1500 &       & 0.556 & 0.219 & \cellcolor[rgb]{ .447,  .765,  .498} 2.55 \\
    Laplace2D2500 &       & 1.478 & 0.550 & \cellcolor[rgb]{ .392,  .749,  .486} 2.69 \\
    Laplace2D4thOdr1000 &       & 0.339 & 0.167 & \cellcolor[rgb]{ .639,  .827,  .537} 2.03 \\
    Laplace2D4thOdr1500 &       & 0.694 & 0.335 & \cellcolor[rgb]{ .624,  .82,  .533} 2.07 \\
    Laplace2D4thOdr2500 &       & 1.857 & 0.871 & \cellcolor[rgb]{ .604,  .816,  .529} 2.13 \\
    Laplace3D100 &       & 0.585 & 0.236 & \cellcolor[rgb]{ .471,  .773,  .502} 2.48 \\
    Laplace3D150 &       & 1.891 & 0.715 & \cellcolor[rgb]{ .408,  .753,  .49} 2.64 \\
    Laplace3D200 &       & 4.459 & 1.718 & \cellcolor[rgb]{ .427,  .757,  .494} 2.60 \\
    Laplace3D250 &       & 8.815 & 3.392 & \cellcolor[rgb]{ .427,  .757,  .49} 2.60 \\
    Laplace3D50 &       & 0.094 & 0.053 & \cellcolor[rgb]{ .741,  .859,  .557} 1.76 \\
    Recirc2D1500 &       & 0.555 & 0.221 & \cellcolor[rgb]{ .459,  .769,  .498} 2.51 \\
    Star2D1000 &       & 0.336 & 0.166 & \cellcolor[rgb]{ .643,  .827,  .537} 2.02 \\
    Star2D1500 &       & 0.715 & 0.331 & \cellcolor[rgb]{ .592,  .812,  .525} 2.16 \\
    Star2D2500 &       & 1.918 & 0.863 & \cellcolor[rgb]{ .569,  .804,  .522} 2.22 \\
    Stretched2D1500 &       & 0.714 & 0.331 & \cellcolor[rgb]{ .596,  .812,  .529} 2.15 \\
    UniFlow2D2500 &       & 1.478 & 0.549 & \cellcolor[rgb]{ .388,  .745,  .482} 2.69 \\
    \end{tabular}%
    } 
  \label{tab:addlabel}%
\end{table*}%



\end{document}